\documentclass[10pt,journal,twocolumn]{IEEEtran}
\usepackage[utf8]{inputenc}
\usepackage{float}
\usepackage{units}
\usepackage{amsmath}
\usepackage{amsthm}
\usepackage{amssymb}
\usepackage{graphicx}
\usepackage[english]{babel}
\usepackage{algorithm}
\usepackage{algorithmic}
\usepackage{amstext}
\usepackage{amsfonts,amssymb}
\usepackage[dvips]{epsfig}
\usepackage{epsfig}
\usepackage[dvips]{graphicx}
\usepackage{multirow}
\usepackage{ltablex}

\theoremstyle{plain}
\newtheorem{thm}{\protect\theoremname}
\theoremstyle{plain}
\newtheorem{lem}[thm]{Lemma}
\theoremstyle{remark}
\newtheorem{rem}[thm]{Remark}
\usepackage{subfig}

\begin{document}

\title{A novel Algorithm for Optimal Placement of Multiple Inertial Sensors
to Improve the Sensing Accuracy}

\author{Nitesh Sahu, Prabhu Babu, Arun Kumar, Rajendar Bahl}
\maketitle
\begin{abstract}
This paper proposes a novel algorithm to determine the optimal placement
of redundant inertial sensors such as accelerometers and gyroscopes
(gyros) for increasing the sensing accuracy. In this paper, we have
proposed a novel iterative algorithm to find the optimal sensor configuration.
The proposed algorithm utilizes the majorization-minimization (MM)
algorithm and the duality principle to find the optimal configuration.
Unlike the state-of-the-art which are mainly geometrical in nature
and restricted to certain noise statistics, the proposed algorithm
gives the exact positions of the sensors, and moreover, the proposed
algorithm is independent of the nature of the noise at different sensors.
The proposed alogrithm has been implemented and tested via numerical
simulation in the MATLAB. The simulation results show that the algorithm
converges to the optimal configurations and show the effectiveness
of the proposed algorithm.
\end{abstract}


\section{Introduction and Literature}

An inertial navigation system (INS) calculates the navigation solution
such as position, velocity, and attitude (PVA) of a navigating platform
based on the initial PVA and by processing the measurements from inertial
measurement unit (IMU) using navigation equations. An IMU is made
of three mutually orthogonal accelerometers and three gyroscopes (gyros)
aligned with the accelerometers. An accelerometer measures the specific
force and gyro measures the angular velocity. The specific force accounts
the acceleration due to all forces except for the gravity \cite{groves2013principles}.
An INS requires at least three accelerometers and three gyros for
computing the navigation solutions; however, using redundant sensors
ensure the reliability and enhance the navigation accuracy. In a redundant
inertial measurement unit (RIMU), more than three gyroscopes or accelerometers
are used to handle the sensor failure or malfunction. When some sensors
malfunction, the faulty sensors should be identified and removed so
that the navigation system with the RIMU continues to operate normally.
Using redundant sensors also enhance the performance of an IMU because
the additional information from the redundant sensors would increase
the sensing accuracy of the IMU \cite{song2016optimal}. However,
using redundant inertial sensors creates the problem of placement
of sensors and orientation of the sensing axis. Here, in this paper
we will investigate the problem of optimal configuration of redundant
sensors to improve the sensing accuracy.

\begin{figure}
\begin{centering}
\subfloat[Class-I configuration for eight sensors.\label{fig:Class-I-config}]{\begin{centering}
\includegraphics[scale=0.46]{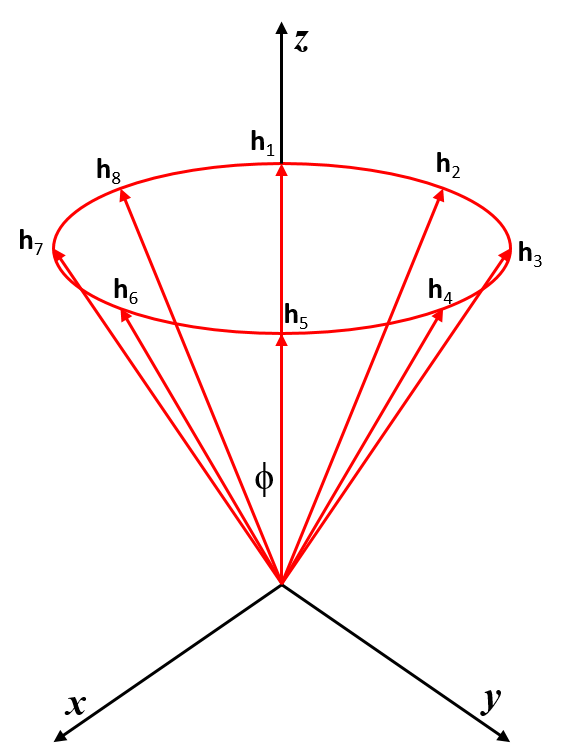}
\par\end{centering}

}\subfloat[Class-II configuration for five sensors.\label{fig:Class-II-config}]{\begin{centering}
\includegraphics[scale=0.46]{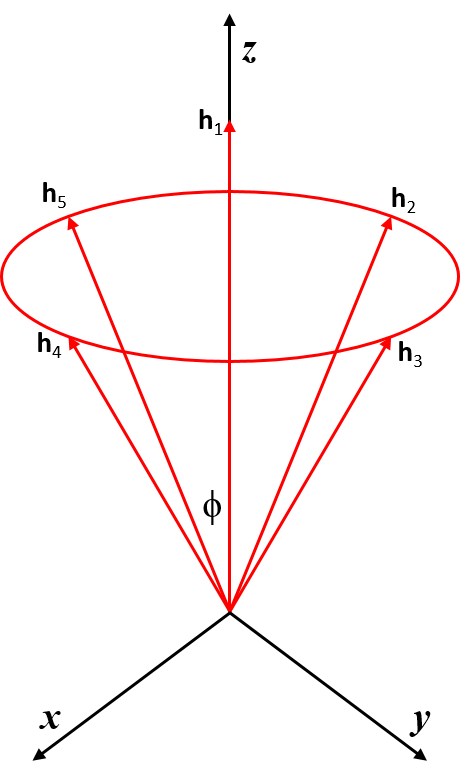}
\par\end{centering}

}
\par\end{centering}
\caption{Class-I and class-II optimal configurations.\label{fig:Class-I-and-class-II}}

\end{figure}

The accuracy analysis of the configuration of sensors having noise
with mean zero and equal variances is described in \cite{pejsa1971optimum}.
In \cite{pejsa1971optimum}, the author has proposed two geometries
for redundant sensor configuration as shown in figure \ref{fig:Class-I-and-class-II}.
In the class-I optimal configuration, the sensors' sensing axis are
equally spaced on a cone with half-angle $\phi=\arccos\left(\nicefrac{1}{\sqrt{3}}\right)$.
Figure \ref{fig:Class-I-config} depicts the class-I configuration
for eight sensors. In the class-II optimal configuration, the sensing
axis of one of the $m$ sensors is placed along cone axis and the
remaining $m-1$ sensors' sensing axis are equally spread on a cone
having half angle $\phi$ satisfying $\cos^{2}\left(\phi\right)=\nicefrac{\left(m-1\right)}{\left(3m-3\right)}$,
where $m$ being the total number of sensors. Figure \ref{fig:Class-II-config}
depicts the class-II configuration for five sensors in which the sensing
axis $\mathbf{h}_{1}$ is placed along the cone axis and the remaining
four sensing axis, $\mathbf{h}_{2}$ to $\mathbf{h}_{5}$, are equally
spread on the cone. In \cite{harrison1977evaluating}, the authors
have presented the configurations of redundant inertial sensors for
improving the navigation performance and for the fault detection and
isolation (FDI) capability; to rank the redundant sensor configuration
for navigation performance, the volume of the ellipsoid associated
with the estimation error covariance matrix, that is, the determinant
of the error covariance matrix has been taken as figure of merit (FOM).
The determinant of the error covariance matrix has also been used
as a FOM in \cite{fu2012novel}, the configuration of sensors has
been proposed considering reliability, navigation accuracy, size and
cost of the system. In \cite{sturza1988skewed}, the configuration
of redundant sensors has been studied for improving sensing accuracy
and detecting faulty sensors. The author in \cite{sturza1988skewed}
has defined the regular polyhedra (platonic solids) and described
the existence of only five regular polyhedra. The platonic solids
have congruent regular polygonal faces and same number of faces meet
at each vertex. It has been suggested that if sensor axes are placed
along the normals to the faces of regular ployhedra they form the
optimal configuration for optimal sensing. In \cite{sturza1988skewed},
geometric dilution of precision (GDOP) which is defined as the square
root of the trace of the estimation error covariance matrix has been
used as FOM to analyze the redundant sensor configurations. The criterion
of minimum GDOP as FOM to analyze the cone configurations has also
been used in \cite{jafari2015optimal}, \cite{jafari2013inertial}.
The idea of maximizing the determinant of the information matrix,
which is the inverse of the estimation error covariance matrix, has
been used in \cite{sukkarieh2000low}, \cite{giroux2004orthogonal}
to determine the optimal configuration of any number of sensors. The
author in \cite{guerrier2009improving} has proposed the partial redundancy
method to determine the optimal configuration of multiple IMUs, which
is based on the concept of reliability and is typically applied in
the geodetic network. In \cite{guerrier2009improving}, it has been
shown that for IMU triads, their optimal configuration is independent
of the geometry between them. The optimal sensor configuration which
considers both the navigation and FDI performances has been discussed
in \cite{shim2010optimal}; a necessary and sufficient condition which
a configuration must satisfy for optimal navigation performance has
been given, and a FOM for sensor configuration which considers both
the navigation and FDI performance has been suggested. The criterion
is that among the optimal sensor configurations for navigation performance,
the optimal configuration for FDI performance is the one which makes
the angle between the nearest two sensors the largest. The authors
in \cite{song2016optimal} have proposed a configuration in which
the orientation of redundant sensors is such that it provides the
best navigation performance and their location minimizes the lever
arm effect. The lever arm effect is generated when the center of gravity
of each sensor deviates from the origin of the case frame. In \cite{song2016optimal},
the maximum eigenvalue of the estimation error covariance matrix has
been taken as the FOM to determine the optimal sensor orientation
for best navigation performance.

\subsection{Our Contribution}

From the literature survey, it was observed that optimal orientations
are discussed for sensors having uncorrelated and equal variances
of random noise. The optimal configurations for such sensors are proposed
to be the platonic solids and sensors placed on a cone (class-I and
class-II configuration) etc. No methods/approaches are available to
optimally place the sensing axis of sensors having correlated and/or
different variances of noise. So one can think of that what would
be the optimal configuration for sensors having different accuracies.
Here, we have proposed a numerical algorithm to find the optimal configuration
of sensors having correlated and different variances of noise. The
proposed algorithm is based on the MM algorithm and the duality theory
from optimization. The proposed algorithm can find the optimal configuration
for any noise configuration. The convergence analysis, computational
complexity and the numerical results of the proposed algorithm are
also discussed.

\subsection{Assumptions and Notations}

In this study, the lever arms are not considered that is their placement
is not considered only orientation has been discussed. The only source
of error in sensor measurement is random noise. The systematic errors
are assumed to be calibrated. The variances of random noise in each
sensor are not the same, and the noise at the sensors are assumed
to be correlated.

Throughout this paper the italic small letters are used for scalars,
boldface small letters for vectors and boldface capital letters for
matrices. $\left(.\right)^{T}$, $\left(.\right)^{-1}$ and $\left\Vert .\right\Vert _{2}$
denote the transpose, inverse and Euclidean norm or $l_{2}-$norm,
respectively. $\mathbb{E}\left(.\right)$, $\det\left(.\right)$,
$\log\left(.\right)$ and $\mathrm{Tr}\left(.\right)$ represent the
expectation, determinant, logarithm and trace operators, respectively.
$\mathbb{R}^{n}$ and $\mathbb{S}_{+}^{n}$denote the $n-$dimensional
Euclidean space and the set of $n\times n$ symmetric positive semidefinite
matrices respectively. Notation $\mathbf{A}\succeq\mathbf{B}$ stands
for matrix $\mathbf{A}-\mathbf{B}$ is positive semidefinite. $\mathbf{x}_{t}$
stands for the value of $\mathbf{x}$ at the $t-$th iteration and
$x_{i}^{t}$ denotes the value of the $i^{th}$ element of $\mathbf{x}_{t}$.

In section \ref{sec:Problem-Formulation}, we discuss the problem
of placing the sensors in optimal configuration and form the optimization
problems whose optimal solution gives the optimal configuration. In
section \ref{sec:ProposedAlgo}, we describe the MM algorithm, the
proposed algorithm which solves the optimization problems formulated
in section \ref{sec:Problem-Formulation}, convergence analysis and
computational complexity. Simulation results are reported in section
\ref{sec:Simulation-Results}. Finally the paper concludes in section
\ref{sec:Conclusion}.

\section{Problem Formulation\label{sec:Problem-Formulation}}

\begin{figure}
\begin{centering}
\includegraphics[scale=0.3]{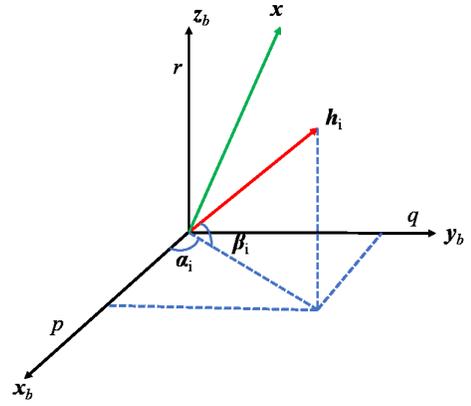}
\par\end{centering}
\caption{Orientation of the single axis inertial sensors\label{fig:OrienSingleAxis}}
\end{figure}

We have the inertial quantity $\mathbf{x}=\left(\begin{array}{ccc}
p & q & r\end{array}\right)^{T}$ which is either acceleration or angular velocity to be measured.
This inertial quantity is to be measured with $m\left(\geq3\right)$
single axis inertial sensors. The gyro and accelerometer measurements
are considered separately. For optimal sensing of the inertial quantity
$\mathbf{x}$, we have to find the optimal configuration of these
$m$ sensors with respect to some reference frame generally taken
as the vehicle body frame ($b-$frame). In figure \ref{fig:OrienSingleAxis},
$x_{b}$, $y_{b}$, and $z_{b}$ are the axes of the $b-$frame.

The unit vector along the sensitive axis of the $i^{th}$ inertial
sensor can be represented in Cartesian coordinate as $\mathbf{h}_{i}\in\mathbb{R}^{n\times1}$
with $n=3$ and $\mathbf{h}_{i}^{T}\mathbf{h}_{i}=1$. 

Let $y_{i}$ denote the component of $\mathbf{x}$ along $\mathbf{h}_{i}$,
the sensitive axis of the $i^{th}$sensor, then $y_{i}$ can be written
as:

\begin{equation}
y_{i}=\mathbf{h}_{i}^{T}\mathbf{x}.
\end{equation}

The actual measurement of $y_{i}$ by the $i^{th}$ sensor with error,
denoted as $\widetilde{y}_{i}$, is given by:

\begin{equation}
\widetilde{y}_{i}=\mathbf{h}_{i}^{T}\mathbf{x}+\varepsilon_{i},
\end{equation}

where $\varepsilon_{i}$ is the zero mean Gaussian random noise with
variance $\sigma_{i}^{2}$, i.e. $\varepsilon_{i}\thicksim N\left(0,\sigma_{i}^{2}\right)$.
The measurements from $m$ redundant inertial sensors can be written
as: 

\begin{equation}
\left(\begin{array}{c}
\widetilde{y}_{1}\\
:\\
:\\
\widetilde{y}_{m}
\end{array}\right)=\left(\begin{array}{c}
\mathbf{h}_{1}^{T}\\
:\\
:\\
\mathbf{h}_{m}^{T}
\end{array}\right)\mathbf{x}+\left(\begin{array}{c}
\varepsilon_{1}\\
:\\
:\\
\varepsilon_{m}
\end{array}\right).
\end{equation}

In vector-matrix form, the above relation can be written as:

\begin{equation}
\widetilde{\mathbf{y}}=\mathbf{H}\mathbf{x}+\boldsymbol{\varepsilon}.
\end{equation}

Here $\widetilde{\mathbf{y}}\in\mathbb{R}^{m\times1}$ is the measurement
vector, $\mathbf{H}\in\mathbb{R}^{m\times n}$ with $n=3$ is the
measurement matrix with $\operatorname{rank}\left(\mathbf{H}\right)=3$,
and $\boldsymbol{\varepsilon}$ is the measurement noise vector which
is Gaussian with mean zero, that is, $\mathbb{E}\left(\boldsymbol{\varepsilon}\right)=0$
and covariance matrix $\mathbf{R}\in\mathbb{S}_{+}^{m}$, given by
$\mathbf{R}=\mathbb{E}\left(\boldsymbol{\varepsilon}\boldsymbol{\varepsilon}^{T}\right)$. 

The weighted least square estimate of $\mathbf{x}$, denoted as $\hat{\mathbf{x}}$,
is given by:

\begin{equation}
\hat{\mathbf{x}}=\left(\mathbf{H}^{T}\mathbf{R}^{-1}\mathbf{H}\right)^{-1}\mathbf{H}^{T}\mathbf{R}^{-1}\mathbf{\widetilde{\mathbf{y}}}.
\end{equation}

The estimation error, denoted as $\mathbf{e}$, is given by:

\begin{equation}
\mathbf{e}=\hat{\mathbf{x}}-\mathbf{x}=\left(\mathbf{H}^{T}\mathbf{R}^{-1}\mathbf{H}\right)^{-1}\mathbf{H}^{T}\mathbf{R}^{-1}\boldsymbol{\varepsilon}.\label{eq:EstErr}
\end{equation}

The estimation error covariance matrix would be:

\begin{equation}
\mathbf{C}_{e}=\mathbb{E}\left(\left(\boldsymbol{\hat{\mathbf{x}}-\mathbf{x}}\right)\left(\boldsymbol{\hat{\mathbf{x}}-\mathbf{x}}\right)^{T}\right)=\left(\mathbf{H}^{T}\mathbf{R}^{-1}\mathbf{H}\right)^{-1}.\label{eq:EstErrCov}
\end{equation}

Since $\boldsymbol{\varepsilon}$ is Gaussian distributed with zero
mean and any linear transformation does not change the Gaussian property,
therefore, the estimation error $\mathbf{e}$, in (\ref{eq:EstErr})
is also zero mean Gaussian distributed, hence, $\mathbf{e}\sim N\left(0,\mathbf{C}_{e}\right)$.
The error covariance matrix characterizes the confidence of the estimation.
The accuracy of the navigation solution of the INS depends on the
error covariance matrix. From (\ref{eq:EstErrCov}), we observe that
the estimation error covariance matrix $\mathbf{C}_{e}$ depends on
sensor configuration matrix $\mathbf{H}$ and the covariance matrix
$\mathbf{R}$ of the measurement noise. Since $\mathbf{R}$ is fixed,
we can choose the $\mathbf{H}$ in such a way that the error covariance
matrix is small in some sense. There are various scalar metrics available
which characterizes the size of the covariance matrix such as determinant,
trace etc. The probability density function for the estimation error,
$\mathbf{e}$, in (\ref{eq:EstErr}) can be written as:

\begin{equation}
f_{\mathbf{e}}\left(\mathbf{e}\right)=\left(2\pi\right)^{-n/2}det\left(\mathbf{C}_{e}\right)^{-1/2}\exp\left(-\frac{1}{2}\mathbf{e}^{T}\mathbf{C}_{e}^{-1}\mathbf{e}\right).
\end{equation}

The equidensity contours of any multivariate Gaussian distribution
form ellipsoid. The locus of points $\mathbf{e}$ defined by $\mathbf{e}^{T}\mathbf{C}_{e}^{-1}\mathbf{e}=k$
forms an ellipsoid, on which the probability density is constant.
For a given $k$ the volume of this ellipsoid is given by \cite{harrison1977evaluating}:

\begin{equation}
V=\frac{4}{3}k^{3/2}\pi\sqrt{\det\left(\mathbf{C}_{e}\right)}.\label{eq:volErrElips}
\end{equation}

From (\ref{eq:volErrElips}) we see that the volume of the error ellipsoid
is proportional to the $\det\left(\mathbf{C}_{e}\right)$. Smaller
the volume of the error ellipsoid, smaller is the estimation error
and better is the estimate obtained using the sensor configuration.
So minimizing the determinant of the error covariance
matrix can be taken as a criterion to find the optimal configuration.

So, to find the optimal configuration of the inertial sensors, we
can minimize the determinant of the error covariance matrix. This
is called the $D-$optimal design \cite{he2010laplacian}. This is
equivalent to minimizing the volume of the resulting confidence ellipsoid.
So in $D-$optimal design we solve the following optimization problem:

\begin{equation}
\begin{aligned} & \underset{\mathbf{H}}{\text{minimize}} &  & \det\left(\left(\mathbf{H}^{T}\mathbf{R}^{-1}\mathbf{H}\right)^{-1}\right)\\
 & \text{subject to} &  & \mathbf{h}_{i}^{T}\mathbf{h}_{i}=1,\;i=1,\ldots,m
\end{aligned}
.\label{eq:dOptimalProb}
\end{equation}

Since logarithm function is monotonic increasing function, taking
the logarithm of the objective function does not change the optimal
solution. Therefore, the problem in (\ref{eq:dOptimalProb}) becomes:

\begin{equation}
\begin{aligned} & \underset{\mathbf{H}}{\text{minimize}} &  & \log\det\left(\left(\mathbf{H}^{T}\mathbf{R}^{-1}\mathbf{H}\right)^{-1}\right)\\
 & \text{subject to} &  & \mathbf{h}_{i}^{T}\mathbf{h}_{i}=1,\;i=1,\ldots,m
\end{aligned}
.\label{eq:OrigMinProb}
\end{equation}

Another criterion that can be used to find the optimal orientation
is to minimize the trace of the error covariance matrix, this is called
the $A-$optimal design \cite{he2010laplacian}. The $A-$optimal
design is also equivalent to minimizing the mean of the squared $l_{2}-$norm
of estimation error $\mathbf{e}$ \cite{boyd2004convex}: 

\begin{equation}
\mathbb{E}\left(\left\Vert \mathbf{e}\right\Vert _{2}^{2}\right)=\mathbb{E}\left(\mathbf{e}^{T}\mathbf{e}\right)=\mathbb{E}\left(\mathrm{Tr}\left(\mathbf{e}\mathbf{e}^{T}\right)\right)=\mathrm{Tr}\left(\mathbf{C}_{e}\right).
\end{equation}

So, in $A-$optimal design, we solve the following optimization problem:

\begin{equation}
\begin{aligned} & \underset{\mathbf{H}}{\text{minimize}} & \mathrm{} & \mathrm{Tr}\left(\left(\mathbf{H}^{T}\mathbf{R}^{-1}\mathbf{H}\right)^{-1}\right)\\
 & \text{subject to} &  & \mathbf{h}_{i}^{T}\mathbf{h}_{i}=1,\;i=1,\ldots,m
\end{aligned}
.\label{eq:OrigMinProb1}
\end{equation}

For the case, when all the sensors have equal accuracy and noises
in them are uncorrelated, we have $\mathbf{R}=\sigma^{2}\mathbf{I}_{m}$,
then the error covariance matrix in (\ref{eq:EstErrCov}) becomes
$\sigma^{2}\left(\mathbf{H}^{T}\mathbf{H}\right)^{-1}$. In this case,
the necessary and sufficient condition which optimal $\mathbf{H}$
must satisfy is \cite{shim2010optimal}:

\begin{equation}
\mathbf{H}_{*}^{T}\mathbf{H}_{*}=\frac{m}{3}\mathbf{I}_{n},\label{eq:OptCondSameSens}
\end{equation}

where $\mathbf{H}_{*}$ is the optimal measurement matrix and $m$
is the number of sensors. 

Next we see what happens when all the sensor axis are rotated by the
same angle about some rotation axis, that is, by the same rotation
matrix. Let the $i^{th}$ sensor axis $\mathbf{h}_{i}$ is rotated
to $\widetilde{\mathbf{h}}_{i}$ by the rotation matrix $\mathbf{C}$.
We can write $\widetilde{\mathbf{h}}_{i}$ as:

\begin{equation}
\widetilde{\mathbf{h}}_{i}=\mathbf{C}\mathbf{h}_{i}\:\forall i=1,\ldots,m.
\end{equation}

Then we have the new configuration matrix $\widetilde{\mathbf{H}}$
whose rows are $\widetilde{\mathbf{h}}_{i}^{T}$. The new rotated
configuration matrix $\widetilde{\mathbf{H}}$ can be written in terms
of $\mathbf{H}$ as follows:

\begin{equation}
\widetilde{\mathbf{H}}=\mathbf{H}\mathbf{C}^{T}.\label{eq:OrigH_RotatedH}
\end{equation}

The rotation matrices are the orthogonal so we have $\mathbf{C}^{T}\mathbf{C}=\mathbf{C}\mathbf{C}^{T}=\mathbf{I}_{n}$.
Using the orthogonal property of the rotation matrix we get the following
relations:
\begin{equation}
\begin{array}{c}
\det\left(\left(\widetilde{\mathbf{H}}^{T}\mathbf{R}^{-1}\widetilde{\mathbf{H}}\right)^{-1}\right)=\det\left(\left(\mathbf{H}^{T}\mathbf{R}^{-1}\mathbf{H}\right)^{-1}\right)\\
\mathrm{Tr}\left(\left(\widetilde{\mathbf{H}}^{T}\mathbf{R}^{-1}\widetilde{\mathbf{H}}\right)^{-1}\right)=\mathrm{Tr}\left(\left(\mathbf{H}^{T}\mathbf{R}^{-1}\mathbf{H}\right)^{-1}\right)
\end{array}.\label{eq:RotatedObjVal}
\end{equation}

From (\ref{eq:RotatedObjVal}) and (\ref{eq:OrigH_RotatedH}), we
conclude that if $\mathbf{H}_{*}$ is the optimal solution for the
problems (\ref{eq:dOptimalProb}) and (\ref{eq:OrigMinProb1}) then
$\mathbf{H}_{*}\mathbf{C}^{T}$ is also the optimal solution. This
means that if a configuration is optimal then the rotated configuration
is also optimal.

In the next section, we describe the proposed algorithm in detail
for solving the problems (\ref{eq:OrigMinProb}) and (\ref{eq:OrigMinProb1}).

\section{Proposed Method\label{sec:ProposedAlgo}}

Before we discuss the proposed algorithm to find the optimal configuration
of the inertial sensors, we first discuss the MM algorithm in section
\ref{subsec:MM-algorithm}, the proposed algorithm to solve the problem
(\ref{eq:OrigMinProb1}) in section \ref{subsec:Primal-Dual-MM_algo}
and in section \ref{subsec:Solving-D-optimal-Problem} we describe
some modification to solve the $D-$optimal problem (\ref{eq:OrigMinProb}).
Finally the convergence analysis and computational complexity of the
proposed algorithm are described in section \ref{subsec:ConverAnalysisProposedAlgo}
and \ref{subsec:Computational-Complexity} respectively.

\subsection{MM algorithm\label{subsec:MM-algorithm}}

The MM algorithm is an iterative method to solve optimization problems
that are difficult to solve directly. The main idea behind this approach
is to convert the difficult objective function into the simpler function
at each iteration. In MM algorithm, at each iteration a surrogate
function which majorizes the actual objective function is minimized.
The MM procedure consists of two steps. In the majorization step,
we find a surrogate function that locally approximates the objective
function at the current iteration point. In other words, the surrogate
function upperbounds the objective function. Then in the minimization
step, we minimize the surrogate function \cite{sun2017majorization}. 

Suppose we want to solve the following constraint optimization problem
iteratively using the MM algorithm:

\begin{equation}
\begin{aligned} & \underset{\mathbf{x}\in\mathcal{X}}{\text{minimize}} &  & f\left(\mathbf{x}\right)\end{aligned}
.
\end{equation}

Suppose $\mathbf{x}_{t}$ is the current iteration point. In the majorization
step, we form a surrogate function $g\left(\mathbf{x}\mid\mathbf{x}_{t}\right)$
at $\mathbf{x}_{t}$, which upperbounds the original objective function
$f\left(\mathbf{x}\right)$. A function $g\left(\mathbf{x}\mid\mathbf{x}_{t}\right)$
is said to majorize a function $f\left(\mathbf{x}\right)$ at $\mathbf{x}_{t}$
if \cite{hunter2004tutorial}

\begin{equation}
f\left(\mathbf{x}_{t}\right)=g\left(\mathbf{x}_{t}\mid\mathbf{x}_{t}\right),\label{eq:SuroFunAndObjFunEq}
\end{equation}

\begin{equation}
f\left(\mathbf{x}\right)\leq g\left(\mathbf{x}\mid\mathbf{x}_{t}\right).\label{eq:MajorizeFunProperty}
\end{equation}

Then, in the minimization step, we update $\mathbf{x}$ as

\begin{equation}
\mathbf{x}_{t+1}\in\text{arg }\underset{\mathbf{x}\in\mathcal{X}}{\text{min }}g\left(\mathbf{x}\mid\mathbf{x}_{t}\right),\label{eq:MM_NextItrPt}
\end{equation}

where $\mathcal{X}$ is the domain of the original optimization problem.
Instead of computing a minimizer of $g\left(\mathbf{x}\mid\mathbf{x}_{t}\right)$,
we can find a point $\mathbf{x}_{t+1}$ that satisfies $g\left(\mathbf{x}_{t+1}\mid\mathbf{x}_{t}\right)\leq g\left(\mathbf{x}_{t}\mid\mathbf{x}_{t}\right)$
\cite{sun2017majorization}.

\begin{figure}
\begin{centering}
\includegraphics[scale=0.22]{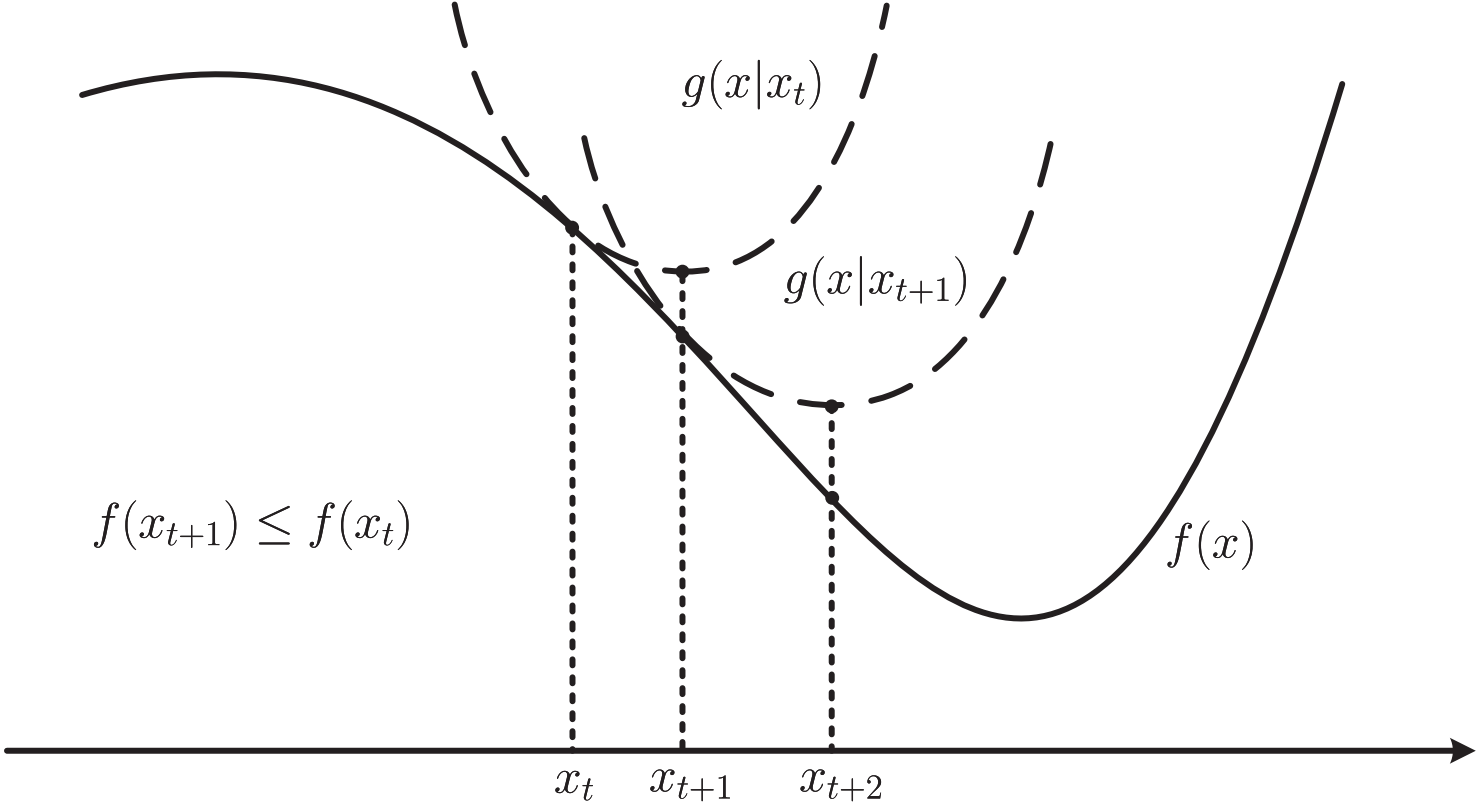}
\par\end{centering}
\caption{The MM procedure \cite{sun2017majorization}.\label{fig:The-MM-procedure}}

\end{figure}

The MM algorithm is described in figure \ref{fig:The-MM-procedure}.
The key feature of the MM algorithm is that at each iteration it decreases
the objective function monotonically

\begin{equation}
f\left(\mathbf{x}_{t+1}\right)\leq g\left(\mathbf{x}_{t+1}\mid\mathbf{x}_{t}\right)\leq g\left(\mathbf{x}_{t}\mid\mathbf{x}_{t}\right)=f\left(\mathbf{x}_{t}\right).\label{eq:MM_DecreasingFeature}
\end{equation}

The first inequality from the right side follows from the definition
of $\mathbf{x}_{t+1}$ in (\ref{eq:MM_NextItrPt}) and second inequality
follows from (\ref{eq:MajorizeFunProperty}).

\subsection{MM over Primal and Dual Variables\label{subsec:Primal-Dual-MM_algo}}

\begin{figure}
\begin{centering}
\includegraphics[scale=0.5]{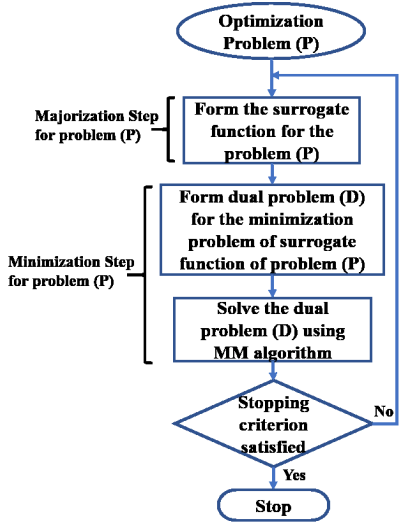}
\par\end{centering}
\caption{Flowchart for MM over primal dual variables.\label{fig:FlowChartPrimalDualMM}}
\end{figure}

In the proposed algorithm, we are utilizing the MM algorithm and the
duality principle from the optimization theory to solve the problem
in (\ref{eq:OrigMinProb1}). The problem in (\ref{eq:OrigMinProb1})
is the original optimization problem which we want to solve. We solve
this problem with MM algorithm, iteratively. Each iteration of the
MM algorithm involves two steps: majorization and minimization. In
the majorization step, we form the surrogate function for the objective
function of the original problem (\ref{eq:OrigMinProb1}) which majorizes
the objective function. In the minimization step, instead of minimizing
the surrogate function, we form the dual problem for the problem of
minimizing the surrogate function, and solve the dual problem using
the MM algorithm. Sometimes after the majorization or minorization
step, the surrogate function which we get is not easy to solve or
the surrogate function can't be solved iteratively via the MM algorithm.
In such cases, we make the dual problem for the minimization problem
of the surrogate function and solve the dual problem over dual variable
by MM algorithm. If strong duality holds then from the dual optimal
solution we can find the primal optimal solution. Therefore, we call
this proposed algorithm as MM over primal dual variable. Figure \ref{fig:FlowChartPrimalDualMM}
shows the flowchart for the proposed algorithm. Now we develop the
idea of MM over primal dual variables for the problem (\ref{eq:OrigMinProb1}).

The optimization problem (\ref{eq:OrigMinProb1}) is

\begin{equation}
\begin{aligned} & \underset{\mathbf{H}}{\text{minimize}} &  & f_{0}\left(\mathbf{H}\right)\triangleq\mathrm{Tr}\left(\left(\mathbf{H}^{T}\mathbf{R}^{-1}\mathbf{H}\right)^{-1}\right)\\
 & \text{subject to} &  & \mathbf{h}_{i}^{T}\mathbf{h}_{i}=1,\;i=1,\ldots,m
\end{aligned}
\label{eq:OrigOptProb}
\end{equation}

where $f_{0}\left(\mathbf{H}\right)$ is the objective function, $\mathbf{H}\in\mathbb{R}^{m\times n}$
is variable, $\mathbf{R}\in\mathbb{S}_{m}^{+}$ is measurement noise
covariance matrix and $\mathbf{h}_{i}^{T}$ is the $i^{th}$ row of
$\mathbf{H}$. Let $D_{f_{0}}$ be the domain of the optimization
problem (\ref{eq:OrigOptProb}) then we can write $D_{f_{0}}$ as

\begin{equation}
D_{f_{0}}=\left\{ \mathbf{H}\in\mathbb{R}^{m\times n}\mid\mathbf{h}_{i}^{T}\mathbf{h}_{i}=1,\;i=1,\ldots,m\right\} \label{eq:DomOrigOptProb}
\end{equation}

The epigraph form for the problem (\ref{eq:OrigOptProb}) is

\begin{equation}
\begin{aligned} & \underset{\mathbf{\mathbf{H}}\in D_{f_{0}},a_{1},\ldots,a_{n}}{\text{minimize}} &  & \sum_{k=1}^{n}a_{k}\\
 & \text{subject to} &  & \left[\begin{array}{cc}
a_{k} & \mathbf{e}_{k}^{T}\\
\mathbf{e}_{k} & \mathbf{H}^{T}\mathbf{R}^{-1}\mathbf{H}
\end{array}\right]\succeq0,\\
 &  &  & k=1,\ldots,n
\end{aligned}
\label{eq:EpiFormOrigOptProb}
\end{equation}

where $\mathbf{e}_{k}$ is the $k^{th}$ column of the $n\times n$
identity matrix $\mathbf{I}_{n}$.

Reformulating problem (\ref{eq:EpiFormOrigOptProb}) we get:

\begin{equation}
\begin{aligned} & \underset{\mathbf{\mathbf{H}}\in D_{f_{0}},a_{1},\ldots,a_{n}}{\text{minimize}} &  & \sum_{k=1}^{n}a_{k}\\
 & \text{subject to} &  & \mathbf{H}^{T}\mathbf{R}^{-1}\mathbf{H}\succ0\\
 &  &  & a_{k}-\mathbf{e}_{k}^{T}\left(\mathbf{H}^{T}\mathbf{R}^{-1}\mathbf{H}\right)^{-1}\mathbf{e}_{k}\succeq0,\\
 &  &  & k=1,\ldots,n
\end{aligned}
.\label{eq:QMI_prog}
\end{equation}

Next we discuss briefly about the difference of convex (DC) programming.
In fact, the DC programming can be shown to be a special case of MM
algorithm. In DC programming, difference of convex function is minimized
with constraints being also the difference of convex functions. The
DC programming can also be seen as the minimization of the sum of
a convex and a concave function with constraints being the sum of
convex and concave functions. The DC programming is not a convex problem
but it can be converted to a convex programming problem by replacing
the concave functions in the objective function and the constraints
by its first order Taylor approximation at current iteration point
\cite{lipp2016variations}. With this substitution the DC programming
becomes a convex problem and can be solved via standard tool \cite{grant2015cvx}.
Similar approach can be adopted for the problem (\ref{eq:QMI_prog})
in which we replace $\mathbf{H}^{T}\mathbf{R}^{-1}\mathbf{H}$ by
$\boldsymbol{\Phi}\left(\mathbf{H};\mathbf{H}_{t}\right)\triangleq-\mathbf{H}_{t}^{T}\mathbf{R}^{-1}\mathbf{H}_{t}+\mathbf{H}^{T}\mathbf{R}^{-1}\mathbf{H}_{t}+\mathbf{H}_{t}^{T}\mathbf{R}^{-1}\mathbf{H}$,
its first order Taylor approximation at $\mathbf{H}_{t}$. With this
subsitution in problem (\ref{eq:QMI_prog}) we get:

\begin{equation}
\begin{aligned} & \underset{\mathbf{\mathbf{H}}\in D_{f_{0}},a_{1},\ldots,a_{n}}{\text{minimize}} &  & \sum_{k=1}^{n}a_{k}\\
 & \text{subject to} &  & \boldsymbol{\Phi}\left(\mathbf{H};\mathbf{H}_{t}\right)\succ0\\
 &  &  & a_{k}-\mathbf{e}_{k}^{T}\left(\boldsymbol{\Phi}\left(\mathbf{H};\mathbf{H}_{t}\right)\right)^{-1}\mathbf{e}_{k}\succeq0,\\
 &  &  & k=1,\ldots,n
\end{aligned}
\label{eq:DC_prog_approach}
\end{equation}

If we undo the epigraph form for problem (\ref{eq:DC_prog_approach})
we get:

\begin{equation}
\begin{aligned} & \underset{\mathbf{\mathbf{H}}\in D_{f_{0}}}{\text{minimize}} &  & \mathrm{Tr}\left(\left(\boldsymbol{\Phi}\left(\mathbf{H};\mathbf{H}_{t}\right)\right)^{-1}\right)\\
 & \text{subject to} &  & \boldsymbol{\Phi}\left(\mathbf{H};\mathbf{H}_{t}\right)\succ0
\end{aligned}
\label{eq:UndoEpi}
\end{equation}

The objective function of the problem (\ref{eq:UndoEpi}) is the surrogate
function for $f_{0}\left(\mathbf{H}\right)$ which upperbounds it.
We Denote the objective function of problem (\ref{eq:UndoEpi}) by:

\begin{equation}
g_{f_{0}}\left(\mathbf{H}\mid\mathbf{H}_{t}\right)\triangleq\mathrm{Tr}\left(\left(\boldsymbol{\Phi}\left(\mathbf{H};\mathbf{H}_{t}\right)\right)^{-1}\right).
\end{equation}

The surrogate function $g_{f_{0}}\left(\mathbf{H}\mid\mathbf{H}_{t}\right)$
satisfies the following property:

\begin{equation}
\begin{aligned}f_{0}\left(\mathbf{\mathbf{H}_{t}}\right)=g_{f_{0}}\left(\mathbf{H}_{t}\mid\mathbf{H}_{t}\right)\\
f_{0}\left(\mathbf{H}\right)\leq g_{f_{0}}\left(\mathbf{H}\mid\mathbf{H}_{t}\right)\: & \textrm{if}\:\boldsymbol{\Phi}\left(\mathbf{H};\mathbf{H}_{t}\right)\succ0
\end{aligned}
\end{equation}

Reformulating the problem (\ref{eq:DC_prog_approach}) we get:

\begin{equation}
\begin{aligned} & \underset{\mathbf{\mathbf{H}}\in D_{f_{0}},a_{1},\ldots,a_{n}}{\text{minimize}} &  & \sum_{k=1}^{n}a_{k}\\
 & \text{subject to} &  & \left[\begin{array}{cc}
a_{k} & \mathbf{e}_{k}^{T}\\
\mathbf{e}_{k} & \boldsymbol{\Phi}\left(\mathbf{H};\mathbf{H}_{t}\right)
\end{array}\right]\succeq0.\\
 &  &  & k=1,\ldots,n
\end{aligned}
\label{eq:ProgToConvertSDP}
\end{equation}

We can relax the constraint $\mathbf{\mathbf{H}}\in D_{f_{0}}$ in
problem (\ref{eq:ProgToConvertSDP}) to make this problem a semidefinite
programming (SDP). We define the following relaxed constraint:

\begin{equation}
D=\left\{ \mathbf{H}\in\mathbb{R}^{m\times n}\mid\mathbf{h}_{i}^{T}\mathbf{h}_{i}\leq1,\;i=1,\ldots,m\right\} .
\end{equation}

If we now let $\mathbf{\mathbf{H}}\in D$ in problem (\ref{eq:ProgToConvertSDP}),
it becomes SDP. This relaxation does not affect the optimal solution
of the problem (\ref{eq:ProgToConvertSDP}) because the objective
function in problem (\ref{eq:ProgToConvertSDP}) is linear and a linear
objective function achieves its optimal solution at the boundary of
the constraint. After relaxing the constraint $\mathbf{\mathbf{H}}\in D_{f_{0}}$
by $\mathbf{\mathbf{H}}\in D$ in problem (\ref{eq:ProgToConvertSDP}),
we get the following SDP:

\begin{equation}
\begin{aligned} & \underset{\mathbf{\mathbf{H}}\in D,a_{1},\ldots,a_{n}}{\text{minimize}} &  & \sum_{k=1}^{n}a_{k}\\
 & \text{subject to} &  & \left[\begin{array}{cc}
a_{k} & \mathbf{e}_{k}^{T}\\
\mathbf{e}_{k} & \boldsymbol{\Phi}\left(\mathbf{H};\mathbf{H}_{t}\right)
\end{array}\right]\succeq0,\\
 &  &  & k=1,\ldots,n
\end{aligned}
\label{eq:FinalPrimalSDP}
\end{equation}

At the optimal solution of the problem (\ref{eq:FinalPrimalSDP})
the constraint $\mathbf{\mathbf{H}}\in D$ will be tight. The problem
in (\ref{eq:FinalPrimalSDP}) is SDP and can be solved to find the
next update of the variable $\mathbf{\mathbf{H}}$ $i.e.$ $\mathbf{H}_{t+1}$.
Solving this SDP may have some issues: we may have to rely on solver,
complexity of the problem increases when the dimension of the variable
$\mathbf{\mathbf{H}}$ increases, for higer dimension of $\mathbf{\mathbf{H}}$,
solvers may take long time and memory issue may also result. Therefore,
directly solving the SDP in (\ref{eq:FinalPrimalSDP}) is not recommended.
So we form the dual problem for SDP in (\ref{eq:FinalPrimalSDP})
and see if the dual problem can be solved easily and efficiently.
As we progress further we will see that using the proposed algorithm
obviates the need of any solver.

Next we form the Lagrangian for problem (\ref{eq:FinalPrimalSDP}).
Let $\mathbf{G}_{k}=\left(\begin{array}{cc}
p_{k} & \mathbf{q}_{k}^{T}\\
\mathbf{q}_{k} & \mathbf{S}_{k}
\end{array}\right)$ be the Lagrange multiplier for each semidefinite constraint. Then
we can write the Lagrangian as

\begin{equation}
\begin{array}{c}
L\left(\mathbf{H},a_{1},\ldots,a_{n},\mathbf{G}_{1},\ldots,\mathbf{G}_{n}\right)=\sum_{k=1}^{n}a_{k}-\\
\sum_{k=1}^{n}\mathrm{Tr}\left(\mathbf{G}_{k}\left(\begin{array}{cc}
a_{k} & \mathbf{e}_{k}^{T}\\
\mathbf{e}_{k} & \boldsymbol{\Phi}\left(\mathbf{H};\mathbf{H}_{t}\right)
\end{array}\right)\right)
\end{array}\label{eq:LagrangeFun}
\end{equation}

With some manipulations, we can write the equation (\ref{eq:LagrangeFun})
as

\begin{equation}
\begin{array}{c}
L\left(\mathbf{H},a_{1},\ldots,a_{n},\mathbf{G}_{1},\ldots,\mathbf{G}_{n}\right)=\sum_{k=1}^{n}\left(1-p_{k}\right)a_{k}-\\
2\mathrm{Tr}\left(\mathbf{R}^{-1}\mathbf{H}_{t}\mathbf{S}\mathbf{H}^{T}\right)+\sum_{k=1}^{n}\mathrm{Tr}\left(\mathbf{G}_{k}\mathbf{D}_{k}\right)
\end{array}\label{eq:FinalLagFun}
\end{equation}

where $\mathbf{D}_{k}=\left(\begin{array}{cc}
0 & -\mathbf{e}_{k}^{T}\\
-\mathbf{e}_{k} & \mathbf{H}_{t}^{T}\mathbf{R}^{-1}\mathbf{H}_{t}
\end{array}\right)$and $\mathbf{S}=\sum_{k=1}^{n}\mathbf{S}_{k}$.

Now we find the Lagrange dual function. The Lagrange dual function
is defined as the infimum of the Lagrangian over primal variables.
The dual function is concave function of the Lagrange multipliers
(also called the dual variables) irrespective of the primal problem
is convex or not \cite{boyd2004convex}. So the dual function is given
by 

\begin{equation}
g\left(\mathbf{G}_{1},\ldots,\mathbf{G}_{n}\right)=\underset{\mathbf{\mathbf{H}}\in D_{f_{0}},a_{1},\ldots,a_{n}}{\text{inf}}L\left(\mathbf{H},a_{1},\ldots,a,\mathbf{G}_{1},\ldots,\mathbf{G}_{n}\right)
\end{equation}

The Lagrangian in (\ref{eq:FinalLagFun}) is bounded below in $a_{k}$
if $p_{k}=1$, so the dual function becomes

\begin{equation}
g\left(\mathbf{G}_{1},\ldots,\mathbf{G}_{n}\right)=\sum_{k=1}^{n}\mathrm{Tr}\left(\mathbf{G}_{k}\mathbf{D}_{k}\right)+\underset{\mathbf{\mathbf{H}}\in D_{f_{0}}}{\text{inf}}-2\mathrm{Tr}\left(\mathbf{R}^{-1}\mathbf{H}_{t}\mathbf{S}\mathbf{H}^{T}\right)
\end{equation}

Assuming $\mathbf{R}^{-1}\mathbf{H}_{t}=\mathbf{C}_{t}=\left(\begin{array}{c}
\left(\widetilde{\mathbf{c}}_{1}^{t}\right)^{T}\\
\vdots\\
\left(\widetilde{\mathbf{c}}_{m}^{t}\right)^{T}
\end{array}\right)$, we have

\begin{equation}
g\left(\mathbf{G}_{1},\ldots,\mathbf{G}_{n}\right)=\sum_{k=1}^{n}\mathrm{Tr}\left(\mathbf{G}_{k}\mathbf{D}_{k}\right)-2\underset{\mathbf{\mathbf{H}}\in D_{f_{0}}}{\text{sup}}\mathrm{Tr}\left(\mathbf{C}_{t}\mathbf{S}\mathbf{H}^{T}\right)\label{eq:DualFun}
\end{equation}

\begin{lem} \label{lem:lemma2}
Given $\mathbf{b}\in\mathbb{R}^{n}$, $\underset{\mathbf{x}^{T}\mathbf{x}=1}{\text{sup}}\mathbf{b}^{T}\mathbf{x}=\left\Vert \mathbf{b}\right\Vert _{2}$
and occurs at $\mathbf{x}=\frac{\mathbf{b}}{\left\Vert \mathbf{b}\right\Vert _{2}}$.

\end{lem}
Now the supremum of the second term in (\ref{eq:DualFun}), using
the lemma \ref{lem:lemma2} is given by

\begin{equation}
\underset{\mathbf{\mathbf{H}}\in D_{f_{0}}}{\text{sup}}\mathrm{Tr}\left(\mathbf{C}_{t}\mathbf{S}\mathbf{H}^{T}\right)=\sum_{i=1}^{m}\left\Vert \mathbf{S}\widetilde{\mathbf{c}}_{i}^{t}\right\Vert _{2}
\end{equation}
and this supremum occurs at 

\begin{equation}
\mathbf{h}_{i}=\frac{\mathbf{S}\widetilde{\mathbf{c}}_{i}^{t}}{\left\Vert \mathbf{S}\widetilde{\mathbf{c}}_{i}^{t}\right\Vert _{2}}.
\end{equation}

so we have the following dual function:

\begin{equation}
g\left(\mathbf{G}_{1},\ldots,\mathbf{G}_{n}\right)=\sum_{k=1}^{n}\mathrm{Tr}\left(\mathbf{G}_{k}\mathbf{D}_{k}\right)-2\sum_{i=1}^{m}\left\Vert \mathbf{S}\widetilde{\mathbf{c}}_{i}^{t}\right\Vert _{2}
\end{equation}

so the dual problem of the problem (\ref{eq:FinalPrimalSDP}) is 

\begin{equation}
\begin{aligned} & \underset{}{\text{maximize}} &  & \sum_{k=1}^{n}\mathrm{Tr}\left(\left(\begin{array}{cc}
1 & \mathbf{q}_{k}^{T}\\
\mathbf{q}_{k} & \mathbf{S}_{k}
\end{array}\right)\mathbf{D}_{k}\right)-2\sum_{i=1}^{m}\left\Vert \mathbf{S}\widetilde{\mathbf{c}}_{i}^{t}\right\Vert _{2}\\
 & \text{subject to} &  & \left(\begin{array}{cc}
1 & \mathbf{q}_{k}^{T}\\
\mathbf{q}_{k} & \mathbf{S}_{k}
\end{array}\right)\succeq0,\;k=1,\ldots,n\\
 &  &  & \mathbf{S}=\sum_{k=1}^{n}\mathbf{S}_{k}
\end{aligned}
\label{eq:MaxDualSDP}
\end{equation}

Since the problem (\ref{eq:FinalPrimalSDP}) is convex, strong duality
holds, that is, optimal value of problem (\ref{eq:FinalPrimalSDP})
equals the optimal value of the problem (\ref{eq:MaxDualSDP}).

Reformulating (\ref{eq:MaxDualSDP}) we get,

\begin{equation}
\begin{aligned} & \underset{}{\text{minimize}} &  & 2\sum_{i=1}^{m}\left\Vert \mathbf{S}\widetilde{\mathbf{c}}_{i}^{t}\right\Vert _{2}-\sum_{k=1}^{n}\mathrm{Tr}\left(\left(\begin{array}{cc}
1 & \mathbf{q}_{k}^{T}\\
\mathbf{q}_{k} & \mathbf{S}_{k}
\end{array}\right)\mathbf{D}_{k}\right)\\
 & \text{subject to} &  & \left(\begin{array}{cc}
1 & \mathbf{q}_{k}^{T}\\
\mathbf{q}_{k} & \mathbf{S}_{k}
\end{array}\right)\succeq0,\;k=1,\ldots,n\\
 &  &  & \mathbf{S}=\sum_{k=1}^{n}\mathbf{S}_{k}
\end{aligned}
\label{eq:MinDualSDP}
\end{equation}

The dual problem of the problem (\ref{eq:FinalPrimalSDP}) is also
SDP and at the optimal solution of the problem (\ref{eq:MinDualSDP})
the linear matrix inequality will be tight, that is, $\mathbf{S}_{k}=\mathbf{q}_{k}\mathbf{q}_{k}^{T}$
for $k=1,\ldots,n$. Solving the dual SDP in (\ref{eq:MinDualSDP})
is also not cheap so we will solve this dual SDP using MM algorithm.
Substituting $\mathbf{S}_{k}=\mathbf{q}_{k}\mathbf{q}_{k}^{T}$ in
problem (\ref{eq:MinDualSDP}) we have $\mathbf{S}=\sum_{k=1}^{n}\mathbf{S}_{k}=\sum_{k=1}^{n}\mathbf{q}_{k}\mathbf{q}_{k}^{T}=\mathbf{Q}\mathbf{Q}^{T}$
where $\mathbf{Q}=\left(\begin{array}{ccc}
\mathbf{q}_{1} & \ldots & \mathbf{q}_{n}\end{array}\right)$, and the following unconstrained optimization problem is achieved:

\begin{equation}
\underset{\mathbf{Q}}{\text{minimize}}\;2\sum_{i=1}^{m}\left\Vert \mathbf{Q}\mathbf{Q}^{T}\widetilde{\mathbf{c}}_{i}^{t}\right\Vert _{2}+\mathrm{Tr}\left(2\mathbf{Q}^{T}-\mathbf{Q}^{T}\mathbf{H}_{t}^{T}\mathbf{R}^{-1}\mathbf{H}_{t}\mathbf{Q}\right)\label{eq:NonCvxUnconsProblem}
\end{equation}

and the objective function is denoted by $\gamma\left(\mathbf{Q}\right)\triangleq2\sum_{i=1}^{m}\left\Vert \mathbf{Q}\mathbf{Q}^{T}\widetilde{\mathbf{c}}_{i}^{t}\right\Vert _{2}+\mathrm{Tr}\left(2\mathbf{Q}^{T}-\mathbf{Q}^{T}\mathbf{H}_{t}^{T}\mathbf{R}^{-1}\mathbf{H}_{t}\mathbf{Q}\right)$.

Reformulating problem (\ref{eq:NonCvxUnconsProblem}), we have the
following:

\begin{equation}
\begin{aligned} & \underset{\mathbf{Q},\beta_{1},\ldots,\beta_{m}}{\text{minimize}} &  & 2\sum_{i=1}^{m}\sqrt{\beta_{i}}+\mathrm{Tr}\left(2\mathbf{Q}^{T}-\mathbf{Q}^{T}\mathbf{H}_{t}^{T}\mathbf{R}^{-1}\mathbf{H}_{t}\mathbf{Q}\right)\\
 & \text{subject to} &  & \beta_{i}=\left\Vert \mathbf{Q}\mathbf{Q}^{T}\widetilde{\mathbf{c}}_{i}^{t}\right\Vert _{2}^{2},\;i=1,\ldots,m.
\end{aligned}
\label{eq:ccvProblem}
\end{equation}

Denote the objective function of the problem (\ref{eq:ccvProblem})
by $\psi\left(\mathbf{Q},\boldsymbol{\beta}\right)\triangleq2\sum_{i=1}^{m}\sqrt{\beta_{i}}+\mathrm{Tr}\left(2\mathbf{Q}^{T}-\mathbf{Q}^{T}\mathbf{H}_{t}^{T}\mathbf{R}^{-1}\mathbf{H}_{t}\mathbf{Q}\right)$,
where $\boldsymbol{\beta}=\left(\begin{array}{ccc}
\beta_{1} & \ldots & \beta_{m}\end{array}\right)^{T}$, $\psi_{1}\left(\boldsymbol{\beta}\right)\triangleq2\sum_{i=1}^{m}\sqrt{\beta_{i}}$,
and $\psi_{2}\left(\mathbf{Q}\right)\triangleq\mathrm{Tr}\left(2\mathbf{Q}^{T}-\mathbf{Q}^{T}\mathbf{H}_{t}^{T}\mathbf{R}^{-1}\mathbf{H}_{t}\mathbf{Q}\right)$.
Therefore, $\psi\left(\mathbf{Q},\boldsymbol{\beta}\right)=\psi_{1}\left(\boldsymbol{\beta}\right)+\psi_{2}\left(\mathbf{Q}\right)$.
\begin{lem}
\label{lem:UpBndSqrtFun}Given $\beta_{i}^{\tau}$, the function $\sqrt{\beta_{i}}$
can be upperbounded as

\begin{equation}
\sqrt{\beta_{i}}\leq\sqrt{\beta_{i}^{\tau}}+\frac{1}{2\sqrt{\beta_{i}^{\tau}}}\left(\beta_{i}-\beta_{i}^{\tau}\right)
\end{equation}

with equality achieved at $\beta_{i}=\beta_{i}^{\tau}$.
\end{lem}
\begin{IEEEproof}
The proof is obvious when we write the first order Taylor expansion
for the function $\sqrt{\beta_{i}}$ at $\beta_{i}^{\tau}$.
\end{IEEEproof}
\begin{lem}
\label{lem:TrUpBnd}
Given $\mathbf{Q}_{\tau}$, the function $w\left(\mathbf{Q}\right)\triangleq\mathrm{Tr}\left(\mathbf{B}\mathbf{Q}^{T}-\mathbf{Q}^{T}\mathbf{A}\mathbf{Q}\right)$,
with $\mathbf{A}\in\mathbb{S}_{+}^{n}$, $\mathbf{B}\in\mathbb{R}^{n\times n}$,
and $\mathbf{Q}\in\mathbb{R}^{n\times n}$, can be upperbounded as 

\begin{equation}
w\left(\mathbf{Q}\right)\leq w\left(\mathbf{Q}_{\tau}\right)+\mathrm{Tr}\left(\left(\mathbf{B}^{T}-2\mathbf{Q}_{\tau}^{T}\mathbf{A}\right)\left(\mathbf{Q}-\mathbf{Q}_{\tau}\right)\right)
\end{equation}

with equality achieved at $\mathbf{Q}=\mathbf{Q}_{\tau}$.
\end{lem}
\begin{IEEEproof}
Writing the first order Taylor expansion for $w\left(\mathbf{Q}\right)$
at $\mathbf{Q}_{\tau}$, we get the above mentioned result.
\end{IEEEproof}
Using lemma \ref{lem:UpBndSqrtFun}, $\psi_{1}\left(\boldsymbol{\beta}\right)$
can be upperbounded at $\boldsymbol{\beta}^{\tau}$ as:

\begin{equation}
\psi_{1}\left(\boldsymbol{\beta}\right)\leq g_{\psi_{1}}\left(\boldsymbol{\beta}\mid\boldsymbol{\beta}^{\tau}\right),
\end{equation}

where $g_{\psi_{1}}\left(\boldsymbol{\beta}\mid\boldsymbol{\beta}^{\tau}\right)\triangleq\sum_{i=1}^{m}\left(\sqrt{\beta_{i}^{\tau}}+\frac{1}{2\sqrt{\beta_{i}^{\tau}}}\left(\beta_{i}-\beta_{i}^{\tau}\right)\right)$.

The function $\psi_{2}\left(\mathbf{Q}\right)$ can be upperbounded
using lemma \ref{lem:TrUpBnd} as:

\begin{equation}
\psi_{2}\left(\mathbf{Q}\right)\leq g_{\psi_{2}}\left(\mathbf{Q}\mid\mathbf{Q}_{\tau}\right),
\end{equation}

where $g_{\psi_{2}}\left(\mathbf{Q}\mid\mathbf{Q}_{\tau}\right)\triangleq\psi_{2}\left(\mathbf{Q}_{\tau}\right)+2\mathrm{Tr}\left(\mathbf{P}_{t,\tau}\left(\mathbf{Q}-\mathbf{Q}_{\tau}\right)\right)$,
and 

\begin{equation}
\mathbf{P}_{t,\tau}=\mathbf{I}_{n}-\mathbf{Q}_{\tau}^{T}\mathbf{H}_{t}^{T}\mathbf{R}^{-1}\mathbf{H}_{t}.\label{eq:EqnFor_P}
\end{equation}

Therefore the surrogate function which upperbounds the $\psi\left(\mathbf{Q},\boldsymbol{\beta}\right)$
can be written as:

\begin{equation}
\psi\left(\mathbf{Q},\boldsymbol{\beta}\right)\leq g_{\psi}\left(\mathbf{Q},\boldsymbol{\beta}\mid\mathbf{Q}_{\tau},\boldsymbol{\beta}^{\tau}\right),
\end{equation}

where $g_{\psi}\left(\mathbf{Q},\boldsymbol{\beta}\mid\mathbf{Q}_{\tau},\boldsymbol{\beta}^{\tau}\right)=g_{\psi_{1}}\left(\boldsymbol{\beta}\mid\boldsymbol{\beta}^{\tau}\right)+g_{\psi_{2}}\left(\mathbf{Q}\mid\mathbf{Q}_{\tau}\right)$.

Since $\gamma\left(\mathbf{Q}\right)=\psi\left(\mathbf{Q},\boldsymbol{\beta}\right)$
with $\beta_{i}=\left\Vert \mathbf{Q}\mathbf{Q}^{T}\widetilde{\mathbf{c}}_{i}^{t}\right\Vert _{2}^{2}$
for $i=1,\ldots,m$, the surrogate function $g_{\gamma}\left(\mathbf{Q}\mid\mathbf{Q}_{\tau}\right)$
which upperbounds the $\gamma\left(\mathbf{Q}\right)$ at $\mathbf{Q}_{\tau}$
can be written as:

\begin{equation}
g_{\gamma}\left(\mathbf{Q}\mid\mathbf{Q}_{\tau}\right)=g_{\psi}\left(\mathbf{Q},\boldsymbol{\beta}\mid\mathbf{Q}_{\tau},\boldsymbol{\beta}^{\tau}\right)\,\text{with}\,\beta_{i}=\left\Vert \mathbf{Q}\mathbf{Q}^{T}\widetilde{\mathbf{c}}_{i}^{t}\right\Vert _{2}^{2}\label{eq:SurrogateFunForUnconsProb}
\end{equation}

The surrogate function in (\ref{eq:SurrogateFunForUnconsProb}) satisfies:

\begin{equation}
\begin{aligned}\gamma\left(\mathbf{\mathbf{Q}_{\tau}}\right)=g_{\gamma}\left(\mathbf{Q}_{\tau}\mid\mathbf{Q}_{\tau}\right)\\
\gamma\left(\mathbf{Q}\right)\leq g_{\gamma}\left(\mathbf{Q}\mid\mathbf{Q}_{\tau}\right)
\end{aligned}
.
\end{equation}

Now we solve the problem (\ref{eq:ccvProblem}) iteratively using
the MM algorithm. Minimizing the surrogate function $g_{\gamma}\left(\mathbf{Q}\mid\mathbf{Q}_{\tau}\right)$
with respect to $\mathbf{Q}$ gives the next iteration point $\mathbf{Q}_{\tau+1}$. 

\begin{equation}
\begin{aligned}\mathbf{Q}_{\tau+1}=\; & \text{arg }\underset{\mathbf{Q}}{\text{min}}\:g_{\gamma}\left(\mathbf{Q}\mid\mathbf{Q}_{\tau}\right)\end{aligned}
\label{eq:updt_prob_Q}
\end{equation}

Leaving the constant terms in $g_{\gamma}\left(\mathbf{Q}\mid\mathbf{Q}_{\tau}\right)$
the problem (\ref{eq:updt_prob_Q}) can be written as:

\begin{equation}
\mathbf{Q}_{\tau+1}=\text{arg }\underset{\mathbf{Q}}{\text{min}}\left(\sum_{i=1}^{m}\frac{\left\Vert \mathbf{Q}\mathbf{Q}^{T}\widetilde{\mathbf{c}}_{i}^{t}\right\Vert _{2}^{2}}{\left\Vert \mathbf{Q}_{\tau}\mathbf{Q}_{\tau}^{T}\widetilde{\mathbf{c}}_{i}^{t}\right\Vert _{2}}+2\mathrm{Tr}\left(\mathbf{Q}\mathbf{P}_{t,\tau}\right)\right).\label{eq:Q_updt_2}
\end{equation}

Problem (\ref{eq:Q_updt_2}) can be rewritten as

\begin{equation}
\mathbf{Q}_{\tau+1}=\text{arg }\underset{\mathbf{Q}}{\text{min}}\left(\mathrm{Tr}\left(\mathbf{F}_{t,\tau}\mathbf{Q}\mathbf{Q}^{T}\mathbf{Q}\mathbf{Q}^{T}\right)+2\mathrm{Tr}\left(\mathbf{Q}\mathbf{P}_{t,\tau}\right)\right)\label{eq:CoordinatDescentPorb}
\end{equation}

where

\begin{equation}
\mathbf{F}_{t,\tau}=\mathbf{C}_{t}^{T}\mathbf{U}_{t,\tau}\mathbf{C}_{t}\label{eq:EqnFor_F}
\end{equation}

with $\mathbf{U}_{t,\tau}=\left(\begin{array}{ccc}
u_{11}^{t,\tau}\\
 & \ddots\\
 &  & u_{mm}^{t,\tau}
\end{array}\right)$ and each $u_{ii}^{t,\tau}=\frac{1}{\left\Vert \mathbf{Q}_{\tau}\mathbf{Q}_{\tau}^{T}\widetilde{\mathbf{c}}_{i}^{t}\right\Vert _{2}}$.

Since no closed form solution available for problem (\ref{eq:CoordinatDescentPorb})
we resort to coordinate descent method to minimize the problem (\ref{eq:CoordinatDescentPorb})
and find the next iteration point $\mathbf{Q}_{\tau+1}$. In the coordinate
descent method, we minimize the objective function iteratively with
respect to the one variable while keeping the rest of the variables
fixed. Let $\mathbf{Q}_{\tau}$ be the value of the variable $\mathbf{Q}$
at $\tau-$th iteration. To do the coordinate descent with respect
to $q_{ij}$, where $q_{ij}$ is the $\left(i,j\right)$ element of
the variable $\mathbf{Q}$, we put $q_{ij}\mathbf{e}_{i}\mathbf{e}_{j}^{T}+\mathbf{Q}_{\tau}^{ij}$
in place of $\mathbf{Q}$ in (\ref{eq:CoordinatDescentPorb}) and
obtain the quartic polynomial in $q_{ij}$. Here $\mathbf{e}_{i}$
denotes the $i^{th}$ column of the $n\times n$ identity matrix $\mathbf{I}_{n}$
and $\mathbf{Q}_{\tau}^{ij}$ denotes the $\tau-$th iteration point
with its $\left(i,j\right)$ element equal to zero. The quartic polynomial
in $q_{ij}$ is given by 

\begin{equation}
r\left(q_{ij}\right)=aq_{ij}^{4}+bq_{ij}^{3}+cq_{ij}^{2}+dq_{ij}+e\label{eq:quarticPoly}
\end{equation}

The coefficients of polynomial (\ref{eq:quarticPoly}) can be computed
using following relations:

\begin{equation}
\begin{aligned}a= & \mathrm{Tr}\left(\mathbf{F}_{t,\tau}\mathbf{K}_{ij}\right)\\
b= & \mathrm{Tr}\left(\mathbf{F}_{t,\tau}\left(\mathbf{K}_{ij}\mathbf{L}_{ij}+\mathbf{L}_{ij}\mathbf{K}_{ij}\right)\right)\\
c= & \mathrm{Tr}\left(\mathbf{F}_{t,\tau}\left(\mathbf{M}_{ij}\mathbf{K}_{ij}+\mathbf{L}_{ij}^{2}+\mathbf{K}_{ij}\mathbf{M}_{ij}\right)\right)\\
d= & \mathrm{Tr}\left(\mathbf{F}_{t,\tau}\left(\mathbf{M}_{ij}\mathbf{L}_{ij}+\mathbf{L}_{ij}\mathbf{M}_{ij}\right)\right)+2\mathrm{Tr}\left(\mathbf{E}_{ij}\mathbf{P}_{t,\tau}\right)\\
e= & \mathrm{Tr}\left(\mathbf{F}_{t,\tau}\mathbf{M}_{ij}^{2}\right)+2\mathrm{Tr}\left(\mathbf{Q}_{\tau}^{ij}\mathbf{P}_{t,\tau}\right)
\end{aligned}
\label{eq:QuaticPolyCoeff}
\end{equation}

where $\mathbf{E}_{ij}=\mathbf{e}_{i}\mathbf{e}_{j}^{T}$, $\mathbf{E}_{ji}=\mathbf{E}_{ij}^{T}$,
$\mathbf{K}_{ij}=\mathbf{E}_{ij}\mathbf{E}_{ji}$, $\mathbf{L}_{ij}=\mathbf{Q}_{\tau}^{ij}\mathbf{E}_{ji}+\mathbf{E}_{ij}\left(\mathbf{Q}_{\tau}^{ij}\right)^{T}$and
$\mathbf{M}_{ij}=\mathbf{Q}_{\tau}^{ij}\left(\mathbf{Q}_{\tau}^{ij}\right)^{T}$.
\\

The reason for using the coordinate descent method is that the dimension
of the $\mathbf{Q}$ is $n\times n$, that is, $3\times3$ so we have
only nine variables for which we have to do the coordinate descent
and this remains fixed irrespective of the number of sensors $m$.
For this reason the coordinate descent method is preferred here.

Each $q_{ij}$ is updated as follows:

\begin{equation}
q_{ij}^{\tau+1}=\text{arg }\underset{q_{ij}}{\text{min}}\:r\left(q_{ij}\right)
\end{equation}

When the coordinate descent for all $q_{ij}$ is completed we get
the next iteration point $\mathbf{Q}_{\tau+1}$. The coordinate descent
method is repeated many times until we get the minimum for the problem
(\ref{eq:NonCvxUnconsProblem}). Let $\mathbf{Q}_{*}$ be the optimal
solution of the problem (\ref{eq:NonCvxUnconsProblem}) obtained after
$N$ cycle of the coordinate descent then $\mathbf{Q}_{*}=\mathbf{Q}_{\tau+N}$.
After this we compute the next update of the variable $\mathbf{H}$
$i.e$ $\mathbf{H}_{t+1}$. From optimal solution $\mathbf{Q}_{*}$,
we compute optimal $\mathbf{S}$ using $\mathbf{S}_{*}=\mathbf{Q}_{*}\mathbf{Q}_{*}^{T}$. 

The $\mathbf{H}_{t}$ is updated to $\mathbf{H}_{t+1}$ by using following
relations:

\begin{equation}
\mathbf{h}_{i}^{t+1}=\frac{\mathbf{S}_{*}\widetilde{\mathbf{c}}_{i}^{t}}{\left\Vert \mathbf{S}_{*}\widetilde{\mathbf{c}}_{i}^{t}\right\Vert _{2}},\label{eq:h_t_1}
\end{equation}

and

\begin{equation}
\mathbf{H}_{t+1}=\left(\begin{array}{c}
\left(\mathbf{h}_{1}^{t+1}\right)^{T}\\
\vdots\\
\left(\mathbf{h}_{m}^{t+1}\right)^{T}
\end{array}\right).\label{eq:H_t_1}
\end{equation}

\begin{figure}
\begin{centering}
\includegraphics[scale=0.4]{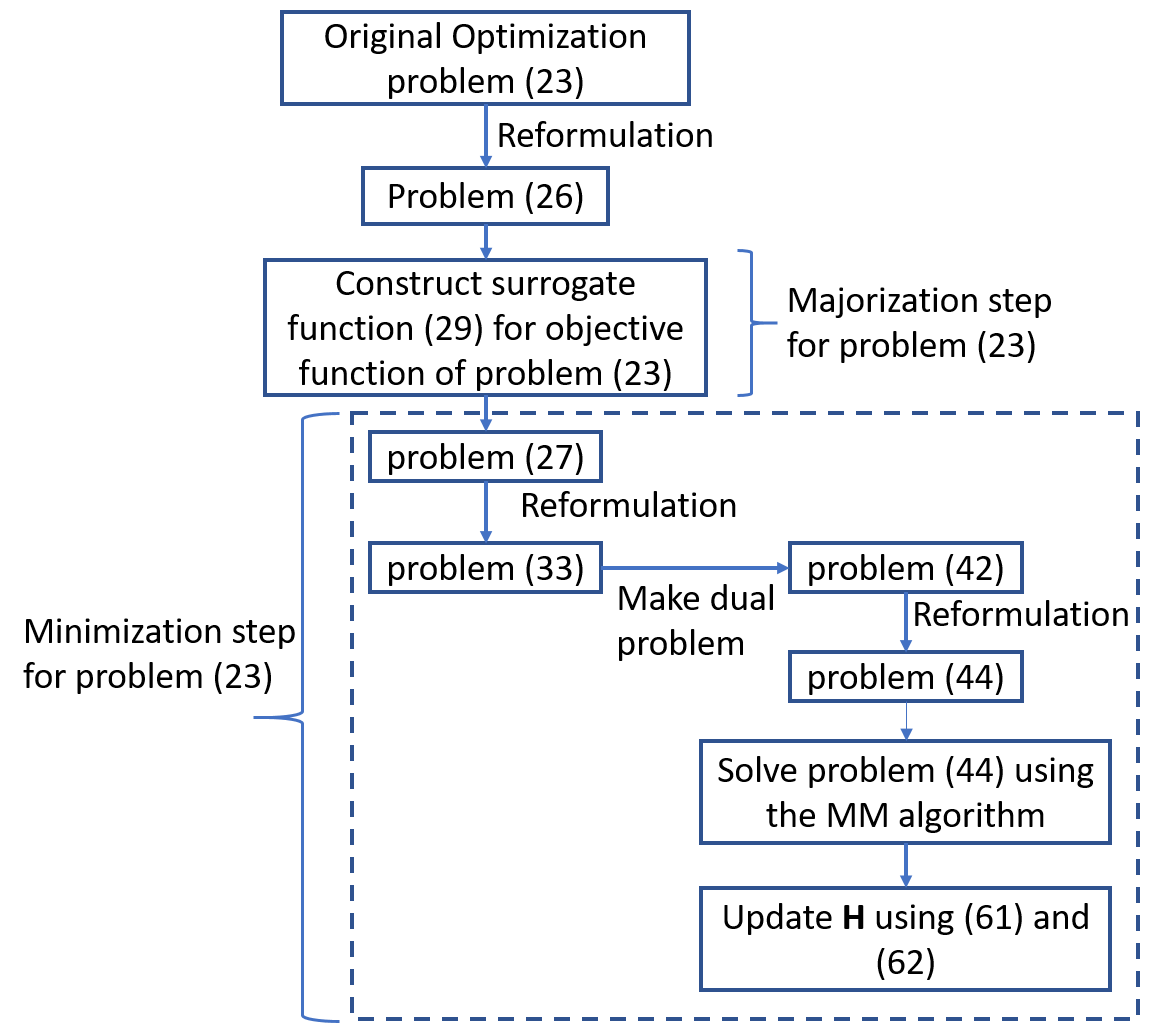}
\par\end{centering}
\caption{Single iteration of the proposed algorithm for problem (\ref{eq:OrigMinProb1}).\label{fig:OneIterProposedAlgo}}
\end{figure}

After obtaining the next iteration $\mathbf{H}_{t+1}$ we repeat the
algorithm multiple times to obtain the optimal solution for the problem
(\ref{eq:OrigOptProb}). Suppose the optimal $\mathbf{H}$ is achieved
after $M$ iteration then $\mathbf{H}_{*}=\mathbf{H}_{t+M}$. The
single iteration of the proposed algorithm is as shown in figure \ref{fig:OneIterProposedAlgo}.
The steps of the proposed primal dual MM algorithm is described in
algorithm \ref{alg:Proposed-primal-dual-MM}.

\begin{algorithm}
\begin{algorithmic}
\STATE Initialize $\mathbf{H}_{0} \in D_{f_{0}}$.\\
\STATE Set $t$, $\tau=0$.\\
{\bf{Repeat}}:
	\STATE \quad Initialize $\mathbf{Q}_{0} \in \mathbb{R}^{n\times n}$. \\
	\quad {\bf{Repeat}}:

		\quad \quad Compute $\mathbf{P}_{t,\tau}$ and $\mathbf{F}_{t,\tau}$ given in \eqref{eq:EqnFor_P} and \eqref{eq:EqnFor_F} respectively.\\
		\quad \quad $\mathbf{Q}_{\tau+1}=\mathbf{Q}_{\tau}$ \\
		\quad \quad {\bf{for}} {$i=1,\ldots,n$} \\
			\quad \quad \quad {\bf{for}} {$j=1,\ldots,n$}
				\STATE \quad \quad \quad \quad Compute $a$, $b$, $c$, $d$ and $e$, the coefficients of the quartic polynomial $r\left(q_{ij}\right)$ in \eqref{eq:quarticPoly} using \eqref{eq:QuaticPolyCoeff}.
				\STATE \quad \quad \quad \quad $q_{ij}^{\tau+1}=\text{arg}\,\underset{q_{ij}}{\text{min}\:}r\left(q_{ij}\right)$
				\STATE \quad \quad \quad \quad $\mathbf{Q}_{\tau+1}\left(i,j\right)=q_{ij}^{\tau+1}$ \\
			\quad \quad \quad {\bf{end}} \\
		\quad \quad {\bf{end}}
	
         \quad $\tau \leftarrow \tau+1$ \\     
	\quad{\bf{until}} convergence
	\STATE Compute $\mathbf{S}_{*}=\mathbf{Q}_{*}\mathbf{Q}_{*}^{T}$, where $\mathbf{Q}_{*}$ is the value of $\mathbf{Q}_{\tau}$ after convergence.
	\STATE Compute $\mathbf{h}_{i}^{t+1}=\frac{\mathbf{S}_{*}\widetilde{\mathbf{c}}_{i}^{t}}{\left\Vert \mathbf{S}_{*}\widetilde{\mathbf{c}}_{i}^{t}\right\Vert _{2}}$ for $i=1,\ldots,m$ and form the $\mathbf{H}_{t+1}$ as described in \eqref{eq:H_t_1}.
\STATE $t \leftarrow t+1$ \\
{\bf{until}} convergence
\STATE {\bf{output}}: $\mathbf{H}_{*}$, where $\mathbf{H}_{*}$ is the value of $\mathbf{H}_{t}$ after convergence.
\end{algorithmic}

\caption{MM algorithm over primal and dual variables.\label{alg:Proposed-primal-dual-MM}}
\end{algorithm}

\subsection{Solving D-optimal Problem by the Proposed Algorithm\label{subsec:Solving-D-optimal-Problem}}

In this section, we solve the problem described in (\ref{eq:OrigMinProb})
using the proposed algorithm. The problem (\ref{eq:OrigMinProb})
can be reformulated as

\begin{equation}
\begin{aligned} & \underset{\mathbf{H},\mathbf{W},\mathbf{B}}{\text{minimize}} &  & \log\det\left(\mathbf{B}\right)\\
 & \text{subject to} &  & \mathbf{h}_{i}^{T}\mathbf{h}_{i}=1,\;i=1,\ldots,m\\
 &  &  & \mathbf{W}=\mathbf{H}^{T}\mathbf{R}^{-1}\mathbf{H}\\
 &  &  & \mathbf{B}=\mathbf{W}^{-1}
\end{aligned}
\label{eq:ReforOrigOptProb}
\end{equation}

The objective function $f\left(\mathbf{B}\right)\triangleq\log\det\left(\mathbf{B}\right)$
in problem (\ref{eq:ReforOrigOptProb}) is concave function and let
$D_{f}$ be the domain of the optimization problem. This concave function
can be minimized iteratively by using the MM algorithm. In the majorization
step, a surrogate function $g_{f}\left(\mathbf{B}\mid\mathbf{B}_{t}\right)$
for the objective function $f\left(\mathbf{B}\right)$ is formed which
upperbounds the objective function at the current point $\mathbf{B}_{t}$. 
\begin{lem}
\label{lem:LogFunUpBound}Given $\mathbf{B}_{t}$, \textup{$\log\det\left(\mathbf{B}\right)$}
can be upper bounded as
\begin{equation}
\log\det\left(\mathbf{B}\right)\leq\log\det\left(\mathbf{B}_{t}\right)+\mathrm{Tr}\left(\mathbf{B}_{t}^{-1}\left(\mathbf{B}-\mathbf{B}_{t}\right)\right)
\end{equation}
with equality achieved at $\mathbf{B}=\mathbf{B}_{t}$ \cite{sun2017majorization}.
\end{lem}
Using lemma \ref{lem:LogFunUpBound},  we get the following surrogate
function at $\mathbf{B}_{t}$ for the objective function of the problem
(\ref{eq:ReforOrigOptProb}):

\begin{equation}
g_{f}\left(\mathbf{B}\mid\mathbf{B}_{t}\right)=\log\det\left(\mathbf{B}_{t}\right)+\mathrm{Tr}\left(\mathbf{B}_{t}^{-1}\left(\mathbf{B}-\mathbf{B}_{t}\right)\right).
\end{equation}

Then in the minimization step, we update $\mathbf{B}$ as

\begin{equation}
\mathbf{B}_{t+1}=\text{arg }\underset{\mathbf{B}\in D_{f}}{\text{min}}\;\mathrm{Tr}\left(\mathbf{B}_{t}^{-1}\mathbf{B}\right)
\end{equation}

The variable $\mathbf{H}$ can be updated as

\begin{equation}
\mathbf{H}_{t+1}=\text{arg }\underset{\mathbf{\mathbf{H}}\in D_{f_{0}}}{\text{min}}\;\mathrm{Tr}\left(\mathbf{W}_{t}\left(\mathbf{H}^{T}\mathbf{R}^{-1}\mathbf{H}\right)^{-1}\right)
\end{equation}

where $\mathbf{W}_{t}=\mathbf{H}_{t}^{T}\mathbf{R}^{-1}\mathbf{H}_{t}$.

So at the current iteration point $\mathbf{H}_{t}$, we have to solve
the following problem to get $\mathbf{H}_{t+1}$:

\begin{equation}
\begin{aligned} & \underset{\mathbf{H}}{\text{minimize}} &  & \mathrm{Tr}\left(\mathbf{W}_{t}\left(\mathbf{H}^{T}\mathbf{R}^{-1}\mathbf{H}\right)^{-1}\right)\\
 & \text{subject to} &  & \mathbf{h}_{i}^{T}\mathbf{h}_{i}=1,\;i=1,\ldots,m
\end{aligned}
\label{eq:PrimalOptProbAftMM}
\end{equation}
Here we observe that problem (\ref{eq:PrimalOptProbAftMM}) is equivalent
to problem (\ref{eq:OrigMinProb1}) if $\mathbf{W}_{t}=\mathbf{I}_{n}$.
Hence we can use the algorithm to solve the $A-$optimal problem (\ref{eq:OrigMinProb1})
iteratively to solve the $D-$optimal problem in (\ref{eq:OrigMinProb}).

\subsection{Convergence Analysis of the Proposed Algorithm\label{subsec:ConverAnalysisProposedAlgo}}

The proposed algorithm is based on the majorization minimization framework
therefore the convergence of the proposed algorithm depends on convergence
of the MM algorithm involved. As described in section \ref{sec:ProposedAlgo},
one MM algorithm is one primal variable $\mathbf{H}$ and other is
on the dual problem with variable $\mathbf{Q}$, so the convergence
will be proved for both the MM algorithms. Since the convergence of
the MM algorithm performed on problem (\ref{eq:OrigOptProb}) depends
on the convergence of the MM performed on problem (\ref{eq:NonCvxUnconsProblem}),
therefore, we will first prove the convergence of the MM algorithm
applied on the problem (\ref{eq:NonCvxUnconsProblem}). The sequence
of points $\left\{ \mathbf{Q}_{\tau}\right\} $ generated by the MM
algorithm monotonically decreases the objective function $\gamma\left(\mathbf{Q}\right)\triangleq2\sum_{i=1}^{m}\left\Vert \mathbf{Q}\mathbf{Q}^{T}\widetilde{\mathbf{c}}_{i}^{t}\right\Vert _{2}+\mathrm{Tr}\left(2\mathbf{Q}^{T}-\mathbf{Q}^{T}\mathbf{H}_{t}^{T}\mathbf{R}^{-1}\mathbf{H}_{t}\mathbf{Q}\right)$
in (\ref{eq:NonCvxUnconsProblem}) as explained in (\ref{eq:MM_DecreasingFeature}).
Moreover function $\gamma\left(\mathbf{Q}\right)$ is bounded below
by zero, since $-\gamma\left(\mathbf{Q}\right)$ is the objective
function of the dual problem of the problem (\ref{eq:FinalPrimalSDP})
whose objective function is bounded below by zero. As strong duality
holds, this implies that $\gamma\left(\mathbf{Q}\right)$ is also
bounded below by zero. Hence the sequence $\left\{ \gamma\left(\mathbf{Q}_{\tau}\right)\right\} $
will converge to some finite value.

A point $\mathbf{Q}$ is called the stationary if:

\begin{equation}
\gamma'\left(\mathbf{Q};\mathbf{D}\right)\geq0,
\end{equation}

where $\gamma'\left(\mathbf{Q};\mathbf{D}\right)$ is the directional
derivative of the matrix function $\gamma\left(\mathbf{Q}\right)$
in the direction of $\mathbf{D}$ and is defined as

\begin{equation}
\gamma'\left(\mathbf{Q};\mathbf{D}\right)=\underset{\alpha\downarrow0}{\lim}\inf\frac{\gamma\left(\mathbf{Q}+\alpha\mathbf{D}\right)-\gamma\left(\mathbf{Q}\right)}{\alpha}.
\end{equation}

From (\ref{eq:MM_DecreasingFeature}), we have

\begin{equation}
\gamma\left(\mathbf{Q}_{0}\right)\geq\gamma\left(\mathbf{Q}_{1}\right)\geq\gamma\left(\mathbf{Q}_{2}\right)\geq\ldots.\label{eq:dualMM_ConverObjFun}
\end{equation}

Assume that there exists a subsequence $\left\{ \mathbf{Q}_{\tau_{j}}\right\} $
which converges to a limit point $\mathbf{Z}$. Then from (\ref{eq:SuroFunAndObjFunEq}),
(\ref{eq:MajorizeFunProperty}) and (\ref{eq:dualMM_ConverObjFun})
we get:

\begin{equation}
\begin{array}{c}
g_{\gamma}\left(\mathbf{Q}_{\tau_{j+1}}\mid\mathbf{Q}_{\tau_{j+1}}\right)=\gamma\left(\mathbf{Q}_{\tau_{j+1}}\right)\leq\gamma\left(\mathbf{Q}_{\tau_{j}+1}\right)\leq\\
g_{\gamma}\left(\mathbf{Q}_{\tau_{j}+1}\mid\mathbf{Q}_{\tau_{j}}\right)\leq g_{\gamma}\left(\mathbf{Q}\mid\mathbf{Q}_{\tau_{j}}\right),
\end{array}\label{eq:ConvAnalEq}
\end{equation}

where $g_{\gamma}\left(.\right)$ is the surrogate function for the
objective function $\gamma\left(\mathbf{Q}\right)$.

Allowing $j\rightarrow\infty$ in (\ref{eq:ConvAnalEq}), we obtain

\begin{equation}
g_{\gamma}\left(\mathbf{Z}\mid\mathbf{Z}\right)\leq g_{\gamma}\left(\mathbf{Q}\mid\mathbf{Z}\right),
\end{equation}

which implies that $g'_{\gamma}\left(\mathbf{Z}\mid\mathbf{Z}\right)\geq0$.
As described in \cite{razaviyayn2013unified}, the first order behavior
of the surrogate function $g_{\gamma}\left(.\right)$ is the same
as the objective function $\gamma\left(.\right)$, so $g'_{\gamma}\left(\mathbf{Z}\mid\mathbf{Z}\right)\geq0$
implies that $\gamma'\left(\mathbf{Z}\right)\geq0$. Hence $\mathbf{Z}$
is the stationary point of $\gamma\left(.\right)$ and hence the MM
algorithm performed on problem (\ref{eq:NonCvxUnconsProblem}) converges
to the stationary point. 

Once the stationay point $\mathbf{Q}_{*}$ is achieved the next iteration
point for $\mathbf{H}$ is computed by using (\ref{eq:h_t_1}) and
(\ref{eq:H_t_1}). Let $\left\{ \mathbf{H}_{t}\right\} $ be the sequence
of points so generated. This sequence of points will monotonically
decrease the objection function of problem (\ref{eq:OrigOptProb}).
Moreover, the objection function of the problem (\ref{eq:OrigOptProb})
is bounded below by zero since the argument of the trace operator
is positive definite. The convergence proof for the MM algorithm performed
on problem (\ref{eq:OrigOptProb}) is the same as the convergence
of MM algorithm performed on problem (\ref{eq:NonCvxUnconsProblem}).
From (\ref{eq:MM_DecreasingFeature}), we have

\begin{equation}
f_{0}\left(\mathbf{H}_{0}\right)\geq f_{0}\left(\mathbf{H}_{1}\right)\geq f_{0}\left(\mathbf{H}_{2}\right)\geq\ldots.\label{eq:PrimallMM_ConverObjFun}
\end{equation}

Assume that there exists a subsequence $\left\{ \mathbf{H}_{t_{j}}\right\} $
converging to the limit point $\mathbf{X}$. Then from (\ref{eq:SuroFunAndObjFunEq}),
(\ref{eq:MajorizeFunProperty}) and (\ref{eq:PrimallMM_ConverObjFun})
we obtain:

\begin{equation}
\begin{aligned}g_{f_{0}}\left(\mathbf{H}_{t_{j+1}}\mid\mathbf{H}_{t_{j+1}}\right)=f_{0}\left(\mathbf{H}_{t_{j+1}}\right)\leq f_{0}\left(\mathbf{H}_{t_{j}+1}\right)\leq\\
g_{f_{0}}\left(\mathbf{H}_{t_{j}+1}\mid\mathbf{H}_{t_{j}}\right)\leq g_{f_{0}}\left(\mathbf{H}\mid\mathbf{H}_{t_{j}}\right),
\end{aligned}
\label{eq:ConvAnalEq-1}
\end{equation}

where $g_{f_{0}}\left(.\right)$ is the surrogate function for the
objective function $f_{0}\left(\mathbf{H}\right)$.

Letting $j\rightarrow\infty$ in (\ref{eq:ConvAnalEq-1}), we get

\begin{equation}
g_{f_{0}}\left(\mathbf{X}\mid\mathbf{X}\right)\leq g_{f_{0}}\left(\mathbf{H}\mid\mathbf{X}\right),
\end{equation}

which implies that $g'_{f_{0}}\left(\mathbf{X}\mid\mathbf{X}\right)\geq0$.
Since the first order behavior of the surrogate function $g_{f_{0}}\left(.\right)$
is the same as the objective function $f_{0}\left(.\right)$ as described
in \cite{razaviyayn2013unified}, so $g'_{f_{0}}\left(\mathbf{X}\mid\mathbf{X}\right)\geq0$
implies that $f_{0}'\left(\mathbf{X}\right)$. Hence point $\mathbf{X}$
is the stationary point of $f_{0}\left(.\right)$ and therefore the
MM algorithm performed on problem (\ref{eq:OrigOptProb}) converges
to the stationary point.

Since both the MM procedure in the proposed algorithm converges, we
conclude that the proposed algorithm converges to the stationary point.

\subsection{Computational Complexity\label{subsec:Computational-Complexity}}

In table \ref{tab:Computational-complexity}, the computational complexity
of the various terms involved in implementing the proposed algorithm,
in terms of number of additions and multiplications, is given. As
the number of sensors, that is, $m$ increases the computational complexity
of $\mathbf{P}_{t,\tau}$, $\mathbf{F}_{t,\tau}$, and $\mathbf{H}_{t+1}$
increases while the computational complexity of the other terms remains
unchanged.

\begin{table}
\begin{centering}
\begin{tabular}{ccc}
\hline 
 & Addition & Multiplication\tabularnewline
\hline 
$\mathbf{P}_{t,\tau}$ & $\mathcal{O}\left(mn^{2}+m^{2}n\right)$ & $\mathcal{O}\left(mn^{2}+m^{2}n\right)$\tabularnewline
$\mathbf{F}_{t,\tau}$ & $\mathcal{O}\left(mn^{3}+m^{2}n\right)$ & $\mathcal{O}\left(mn^{3}+m^{2}n\right)$\tabularnewline
$a$ & $\mathcal{O}\left(n^{3}\right)$ & $\mathcal{O}\left(n^{3}\right)$\tabularnewline
$b$ & $\mathcal{O}\left(n^{3}\right)$ & $\mathcal{O}\left(n^{3}\right)$\tabularnewline
$c$ & $\mathcal{O}\left(n^{3}\right)$ & $\mathcal{O}\left(n^{3}\right)$\tabularnewline
$d$ & $\mathcal{O}\left(n^{3}\right)$ & $\mathcal{O}\left(n^{3}\right)$\tabularnewline
$e$ & $\mathcal{O}\left(n^{3}\right)$ & $\mathcal{O}\left(n^{3}\right)$\tabularnewline
$\mathbf{H}_{t+1}$ & $\mathcal{O}\left(mn^{3}\right)$ & $\mathcal{O}\left(mn^{3}\right)$\tabularnewline
\hline 
\end{tabular}
\par\end{centering}
\caption{Computational complexity.\label{tab:Computational-complexity}}

\end{table}

\section{Simulation Results\label{sec:Simulation-Results}}

The proposed algorithm to find the optimal configuration of the inertial
sensors has been described in section \ref{sec:ProposedAlgo}. The
algorithm has been implemented in MATLAB. In this section, we show
some simulation results which demonstrate that the proposed algorithm
converges to optimal solutions and computes the optimal configuration
for optimal sensing. First, in section \ref{subsec:SensWithSameAccuUncorrNoise},
we will show the optimal configuration for two, three, and four sensors
whose optimal configurations are already discussed in the literature.
In section \ref{subsec:SensWithDiffAccuUncorr}, we will show the
optimal sensor configuration for sensors having different accuracies
and uncorrelated noise and finally in section \ref{subsec:SensWithCorrNoise}
for sensors with correlated noise.
\begin{rem}
This algorithm does not depend on the initialization. The initialization
point is randomly selected in $D_{f_{0}}$ defined in (\ref{eq:DomOrigOptProb}).
We have tried many different initialization points and it is observed
that the proposed algorithm always converges to the same optimal objective
value (optimal solutions may be different) irrespective of the initialization
point. The objective function is not convex but it seems to be a unimodal
and we do not have any mathematical proof for this.
\end{rem}

\subsection{Sensors with Same Accuracies and Uncorrelated Noise\label{subsec:SensWithSameAccuUncorrNoise}}

In this subsection, we consider the optimal configuration for sensors
having same accuracies and uncorrelated noise. In this case, the measurement
noise covariance marix will be some scalar multiple of the identity
matrix. The optimal configuration for such sensors are already discussed
in the literature. We will compute these configuration from the proposed
algorithm.

\subsubsection{Three Sensors in Three Dimensions}

Three sensors with the same accuracy form the optimal configuration
in three-dimensional space when they are orthogonal to each other
regardless of individual sensor's orientation \cite{guerrier2008integration}.
This will be verified by the proposed algorithm. An infinite number
of optimal configurations are possible for three sensors in three-dimensional
space which is orthogonal and the proposed algorithm may converge
to one of them.

For three sensors in three-dimensional space, we have $m=3$, $n=3$,
and take $\mathbf{R}=3\mathbf{I}_{3}$. The one of the optimal $\mathbf{H}$
obtained from the proposed algorithm is

\begin{equation}
\mathbf{H}_{*}=\left(\begin{array}{ccc}
0.2239 & 0.2300 & -0.9471\\
-0.3311 & 0.9319 & 0.1481\\
-0.9166 & -0.2804 & -0.2848
\end{array}\right)
\end{equation}

Here rows of $\mathbf{H}_{*}$ are mutually orthogonal hence verifying
the optimality criterion of three sensors having equal variances of
noise. Figure \ref{fig:ObjFunVal3SensSame} shows the objective function
values plotted at each iteration, we observe that the value of the
objective function decreases at each iteration and converges to the
value $3.295836$ and the optimal solution $\mathbf{H}_{*}$ obtained
from the proposed algorithm satisfies the condition in (\ref{eq:OptCondSameSens}).
This shows that the proposed algorithm converges and computes the
optimal configuration. The optimal configuration obtained is plotted
in figure \ref{fig:Config3SensSame}.

\subsubsection{Four Sensors in Three Dimensions}

In this section, we consider the four sensors with same accuracies
and compute the optimal configurations by the proposed algorithm.
In the literature, three possible optimal configurations have been
proposed for four sensors which are class-I, class-II, and tetrad
configuration \cite{guerrier2008integration}. The proposed algorithm
may converge to any one of these optimal configurations depending
on the initialization, but they may have different orientations.

\begin{figure}
\subfloat[Three sensors in three dimension\label{fig:ObjFunVal3SensSame}]{\begin{centering}
\includegraphics[scale=0.33]{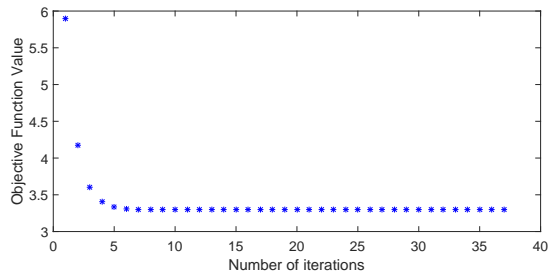}
\par\end{centering}
}

\subfloat[Four sensors in three dimension\label{fig:ObjFunVal4SensSame}]{\begin{centering}
\includegraphics[scale=0.33]{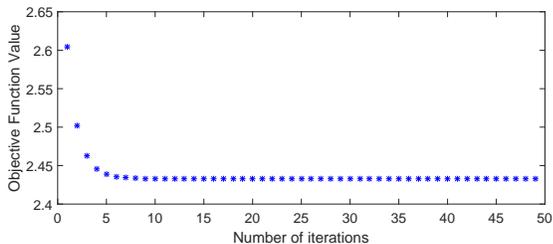}
\par\end{centering}

}
\begin{centering}
\caption{Value of the objective function of problem (\ref{eq:OrigMinProb})
plotted against the number of iterations for sensors with equal accuracies.\label{fig:SimResltFourSensSameVar}}
\par\end{centering}
\end{figure}

\begin{figure}
\subfloat[Three sensors\label{fig:Config3SensSame}]{\begin{centering}
\includegraphics[scale=0.5]{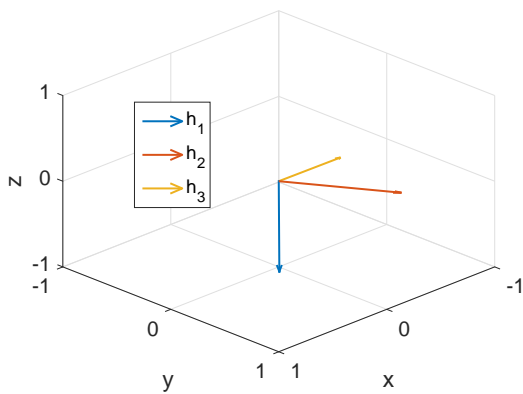}
\par\end{centering}
}

\subfloat[Four sensors\label{fig:Config4SensSame}]{\begin{centering}
\includegraphics[scale=0.5]{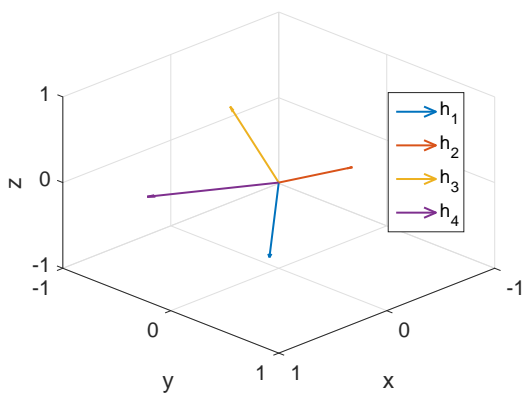}
\par\end{centering}
}

\caption{Optimal configurations obtained by the proposed algorithm for simulation
setup discussed in section \ref{subsec:SensWithSameAccuUncorrNoise}.}

\end{figure}

We have $m=4$, $n=3$, and take $\mathbf{R}=3\mathbf{I}_{4}$. Figure
\ref{fig:ObjFunVal4SensSame} shows the objective function values
plotted at each iteration, we observe that the value of the objective
function decreases at each iteration and converges to the value $2.432790$
and the optimal $\mathbf{H}$ obtained by the proposed algorithm satisfies
the condition mentioned in (\ref{eq:OptCondSameSens}). This shows
that the proposed algorithm converges and computes the optimal configuration.
The optimal configuration obtained is plotted in figure \ref{fig:Config4SensSame}.

\subsection{Sensors with Different Accuracies and Uncorrelated Noise\label{subsec:SensWithDiffAccuUncorr}}

In this section, we show some results for the case when the sensors
have different accuracies and uncorrelated noise, that is, the measurement
noise covariance matrix is diagonal with positive diagonal elements.
The proposed algorithm converges to values mentioned in the table
\ref{tab:ObjFunValDiffAccu} and hence converges to the optimal solutions.

\begin{table}
\begin{centering}
\includegraphics[scale=0.75]{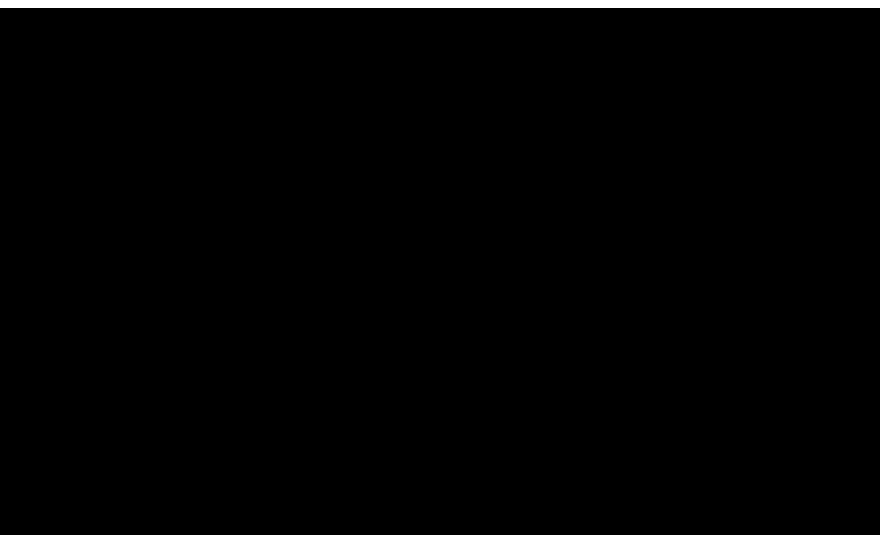}\caption{The objective function values at which the proposed algorithm converges
for diagonal covariance matrices.\label{tab:ObjFunValDiffAccu}}
\par\end{centering}
\end{table}

\subsection{Sensors with correlated noise\label{subsec:SensWithCorrNoise}}

In this section, we show the simulation result for the case when sensors
have correlated noise, that is, the off-diagonal elements of the measurement
noise covariance matrix, $\mathbf{R}$, are non-zero. We randomly
choose $\mathbf{R}$ in $\mathbb{S}_{++}^{m}$, where $\mathbb{S}_{++}^{m}$
represents the set of positive definite matrices of dimension $m\times m$. 

\begin{figure}
\subfloat[The objective function of problem (\ref{eq:OrigMinProb}) plotted
against the number of iterations.\label{fig:ObjFunVal_5SensGen}]{\begin{centering}
\includegraphics[scale=0.33]{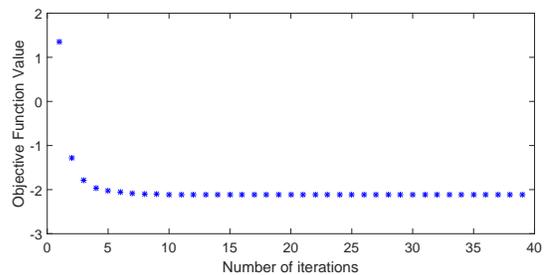}
\par\end{centering}

}

\subfloat[Optimal configurations obtained by the proposed algorithm.\label{fig:OptConfig5SensGen}]{\begin{centering}
\includegraphics[scale=0.45]{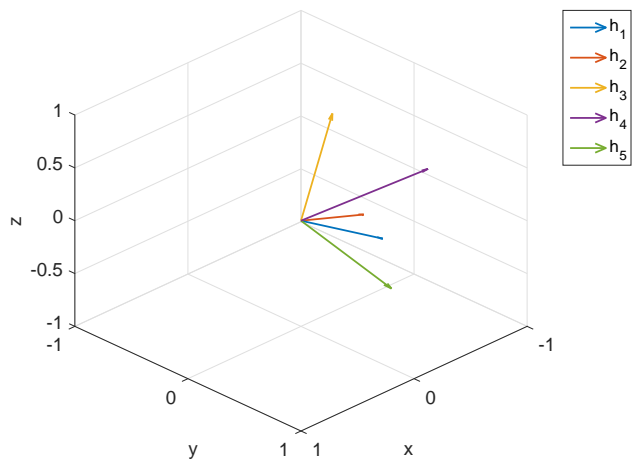}
\par\end{centering}

}

\caption{Simulation results for sensors with correlated noises.}

\end{figure}

We take $m=5$, $n=3$, and choose $\mathbf{R}$ in $\mathbb{S}_{++}^{m}$.
In figure \ref{fig:ObjFunVal_5SensGen}, the values of the objective
function of problem (\ref{eq:OrigMinProb}) are plotted at each iteration.
The value of the objective function decreases at each iteration and
converges to a finite value and a optimal solution is achieved. The
optimal configuration obtained by the proposed algorithm is shown
in figure \ref{fig:OptConfig5SensGen}.

\section{Conclusion\label{sec:Conclusion}}

In this paper, a novel optimization algorithm which computes the optimal
configuration for inertial sensors has been proposed. The proposed
algorithm is based on the MM algorithm and the duality principle from
the optimization theory. The proposed algorithm not only computes
the optimal configuration for sensors with same accuracies but also
for sensors with different accuracies and even for sensors having
the correlated measurement noise.

In literature, optimal configurations are discussed and computed for
sensors having same accuracies and measurement noises in sensors are
uncorrelated. Here, we have extended the idea of optimal configuration
for sensors having different accuracies and correlated noise.

The effectiveness of the proposed algorithm is verified via simulation
results. The results show that the algorithm converges to the optimal
solutions.

\bibliographystyle{ieeetr}
\bibliography{LyxFilePrabhuSirAfterComment8}

\end{document}